\documentclass[twocolumn,journal]{IEEEtran}
\usepackage[T1]{fontenc}
\usepackage{amsthm}
\usepackage{amsmath}
\usepackage{amssymb}
\usepackage{graphicx}
\usepackage[unicode=true,
 bookmarks=true,bookmarksnumbered=true,bookmarksopen=true,bookmarksopenlevel=1,
 breaklinks=false,pdfborder={0 0 0},backref=false,colorlinks=false]
 {hyperref}
\hypersetup{pdftitle={Your Title},
 pdfauthor={Your Name},
 pdfpagelayout=OneColumn, pdfnewwindow=true, pdfstartview=XYZ, plainpages=false}
\usepackage{breakurl}

\makeatletter
  \theoremstyle{definition}
  \newtheorem{defn}{\protect\definitionname}
  \theoremstyle{plain}
  \newtheorem{prop}{\protect\propositionname}
  \theoremstyle{remark}
  \newtheorem{rem}{\protect\remarkname}
  \theoremstyle{definition}
  \newtheorem{example}{\protect\examplename}

\usepackage[vlined,boxed,ruled,linesnumbered,resetcount]{algorithm2e}
\usepackage{algpseudocode}

\makeatother

\providecommand{\definitionname}{Definition}
\providecommand{\examplename}{Example}
\providecommand{\propositionname}{Proposition}
\providecommand{\remarkname}{Remark}

\begin{document}

\title{Boosted KZ and LLL Algorithms}

\author{Shanxiang Lyu and~Cong Ling,~\IEEEmembership{Member,~IEEE}\thanks{This work was supported in part by the Royal Society and in part by
the China Scholarship Council.}\thanks{S. Lyu and C. Ling are with the Department of Electrical and Electronic
Engineering, Imperial College London, London SW7 2AZ, United Kingdom
(e-mail: s.lyu14@imperial.ac.uk, cling@ieee.org). }}
\maketitle
\begin{abstract}
There exist two issues among popular lattice reduction (LR) algorithms
that should cause our concern. The first one is Korkine-Zolotarev
(KZ) and Lenstra\textendash Lenstra\textendash Lovász (LLL) algorithms
may increase the lengths of basis vectors. The other is KZ reduction
suffers much worse performance than Minkowski reduction in terms of
providing short basis vectors, despite its superior theoretical upper
bounds. To address these limitations, we improve the size reduction
steps in KZ and LLL to set up two new efficient algorithms, referred
to as boosted KZ and LLL, for solving the shortest basis problem (SBP)
with exponential and polynomial complexity, respectively. Both of
them offer better actual performance than their classic counterparts,
and the performance bounds for KZ are also improved. We apply them
to designing integer-forcing (IF) linear receivers for multi-input
multi-output (MIMO) communications. Our simulations confirm their
rate and complexity advantages.\end{abstract}


\section{Introduction}

\noindent \IEEEPARstart{L}{attice } reduction (LR) is a process that,
given a lattice basis as input, to ascertain another basis with short
and nearly orthogonal vectors \cite{Micciancio2002}. Their applications
in signal processing include global positioning system (GPS) \cite{Hassibi1998},
color space estimation in JPEG images \cite{Neel2001}, and data detection/precoding
in wireless communications \cite{Yao2002,Windpassinger2004}. Recent
advances in LR algorithms are mostly made in wireless communications
and cryptography \cite{Wubben2011,Nguyen2010}. Popular LR algorithms
with exponential complexity include Korkine-Zolotarev (KZ) \cite{Korkinge1877,Wen2015}
and Minkowski reductions \cite{Zhang2012tsp}, which have set the
benchmarks for the best possible performance in LR aided successive
interference cancellation (SIC) and zero-forcing (ZF) detectors for
multi-input multi-output (MIMO) systems \cite{Zhang2012tsp}. In MIMO
detection problems, KZ and Minkowski reductions are preferable when
the channel coefficients stay fixed for a long time frame so that
their high complexity can be shared across time. In the part of polynomial
or fixed complexity algorithms, the celebrated Lenstra\textendash Lenstra\textendash Lovász
(LLL) \cite{Lenstra1982} algorithm has been well studied and many
new variants have been proposed. Typical variants of LLL in wireless
communications can be summarized into two types: either sacrificing
the execution of full size reductions \cite{Ling2010,Zhang2012},
or controlling the implementation order of swaps and size reductions
\cite{Vetter2009,Wen2014b,Chang2005,Wen2016}. The reason to establish
the first type variants is that a full size reduction has little influence
on the performance of LR aided SIC detectors. Variants of the second
type, e.g., fixed complexity LLL \cite{Vetter2009,Wen2014b} and greedy
LLL \cite{Chang2005,Wen2016}, serve the purpose of enhancing the
system performance especially when the number of LLL iterations is
restrained. It is also noteworthy to introduce the block KZ (BKZ)
reduction \cite{Schnorr1994} as a tradeoff between KZ and LLL. BKZ
is scarcely probed in MIMO but more often in cryptography. Many records
in the shortest vector problem (SVP) challenge hall of fame \cite{Schneider2010}
are set by using BKZ although no good upper bound on the complexity
of BKZ is known. 

In this work, we point out two issues among popular LR algorithms
which were rarely investigated before. The first one is that KZ and
LLL may elongate basis vectors. This issue was discovered when we
applied LLL to Gaussian random matrices of dimensions higher than
$40$. The second one is KZ reduction practically suffers much worse
performance than Minkowski reduction in terms of providing short basis
vectors, while Nguyen and Stehle conjectured in \cite[P. 46:7]{Nguyen2012b}
that KZ may be stronger than Minkowski in high dimensions because
theoretically all vectors of a KZ reduced basis are known to be closer
to the successive minima than Minkowski's. So engineers may be quite
confused about the discrepancies between theory and practice. 

\textit{The contributions of this work are twofold. First, we propose
improved algorithms to address the above limitations of KZ and LLL,
and they are in essence suitable for any application that needs to
solve the shortest basis problem (SBP). Second,} \textit{we show that
our algorithms can be applied to the design of integer forcing (IF)
linear MIMO receivers \cite{Zhan2014IT} to obtain some gains in rates. }

The first algorithm is referred to as boosted KZ. It harnesses the
strongest length reduction every time after the shortest vector in
a projected lattice basis has been found, and such an operation is
proved to be valid. We improve analysis on the best known bounds for
the lengths of basis vectors and Gram-Schmidt vectors via boosted
KZ. After choosing sphere decoding as subroutines, the total complexity
of boosted KZ is shown to be closed to that of conventional KZ.

In the second algorithm called boosted LLL, it also dumps conventional
size reduction conditions in LR while deploying a flexible effort
to perform length reduction. In order to maintain the Siegel condition
\cite{Gama2006}, two criteria for doing length reductions before/after
testing the necessity of swaps are proposed, which guarantee the basis
potential is decreasing after swaps and the lengths of vectors shrink
at the largest extent. With our scheme, bounds on basis lengths and
orthogonal defects can also be obtained. An optimal principle of choosing
Lovász constants is proposed as well. The complexity of this algorithm
is of $O(Ln^{4+c}\ln n)$ if the condition number of the input basis
is of $O(\ln n)$, where $L$ is the number of routes in boosted LLL
and $c>1$ is a constant.

IF is a new MIMO receiver architecture that attempts to decode an
integer combination of lattice codes \cite{Zhan2014IT}. It can be
thought of as a special case of compute and forward \cite{Nazer2011}
because this design has full cooperation among receive antennas. \textit{We
apply our algorithms to IF because it represents the kind of applications
that need to find multiple short lattice vectors, as opposed to LR
aided SIC receivers \cite{Ling2011}, lattice Gaussian samplers \cite{Liu2011it},
and those searching the shortest or closest vectors \cite{Agrell2002,Hassibi2005}.}
This receiver is more general than LR aided minimum mean square error
(MMSE) receiver in that it allows concise evaluation on rates, owning
to lattice coding and dithering. In \cite{Zhan2014IT}, the performance
of IF receiver is shown to outperform conventional ZF and MMSE receivers,
and the optimality in diversity-multiplexing gain tradeoff (DMT) is
also proved. We will elaborate on the IF architecture and the SBP
interface where boosted KZ and LLL turn out to be beneficial. Simulations
will verify the advantages of our algorithms in terms of rates and
complexity.

The rest of this paper is organized as follows. Backgrounds about
lattices and lattice reduction algorithms are reviewed in Section
II. After that, we provide a motivating example to indicate the drawback
of KZ and LLL. The boosted KZ and LLL algorithms are subsequently
constructed and analyzed in Sections III and IV, respectively. After
introducing the IF framework, exemplary simulation results are then
shown in Section V to emphasize that the proposed algorithms can deliver
higher rates. We mention some open questions Section VI. 

Notation: Matrices and column vectors are denoted by uppercase and
lowercase boldface letters. For a matrix $\mathbf{D}$, $\mathbf{D}_{i:j,i:j}$
denotes the submatrix of $\mathbf{D}$ formed by rows and columns
$i,\thinspace i+1,\ldots\thinspace,j$. When referring to the $(i,\thinspace j)$th
element of $\mathbf{D}$, we simply write $d_{i,j}$. $\mathbf{I}_{n}$
and $\mathbf{0}_{n}$ denote the $n\times n$ identity matrix and
$n\times1$ zero vector, and the operation $(\cdot)^{\top}$denotes
transposition. For an index set $\Gamma_{i}=\left\{ 1,\ldots\thinspace,i-1\right\} $,
$\mathbf{D}_{\Gamma_{i}}$ denotes the columns of $\mathbf{D}$ indexed
by $\Gamma_{i}$. $\mathrm{span}(\mathbf{D}_{\Gamma_{i}})$ denotes
the vector space spanned by vectors in $\mathbf{D}_{\Gamma_{i}}$.
$\pi_{\mathbf{D}_{\Gamma_{i}}}(\mathbf{x})$ and $\pi_{\mathbf{D}_{\Gamma_{i}}}^{\bot}(\mathbf{x})$
denote the projection of $\mathbf{x}$ onto $\mathrm{span}(\mathbf{D}_{\Gamma_{i}})$
and the orthogonal complement of $\mathrm{span}(\mathbf{D}_{\Gamma_{i}})$.
$\lfloor x\rceil$ denotes rounding $x$ to the nearest integer, $|x|$
denotes getting the absolute value of $x$, and $\left\Vert \mathbf{x}\right\Vert $
denote the Euclidean norm of vector $\mathbf{x}$. The set of all
$n\times n$ matrices with determinant $\pm1$ and integer coefficients
will be denoted by $\mathrm{GL}_{n}(\mathbb{Z})$.

\section{Preliminaries}

\subsection{Lattices }

A full rank $n$-dimensional lattice $\mathcal{L}$ is a discrete
additive subgroup in $\mathbb{R}^{n}$. The lattice generated by a
basis $\mathbf{D}=[\mathbf{d}_{1},\ldots\thinspace,\mathbf{d}_{n}]\in\mathbb{R}^{n\times n}$
can be written as 
\[
\mathcal{L}(\mathbf{D})=\left\{ \mathbf{v}\mid\mathbf{v}=\sum_{i\in[n]}c_{i}\mathbf{d}_{i}\thinspace;\thinspace c_{i}\in\mathbb{Z}\right\} ;
\]
its dual lattice $\mathcal{L}(\tilde{\mathbf{D}})$ has a basis $\tilde{\mathbf{D}}=\mathbf{D}^{-\top}$.
If the lattice basis is clear from the context, we omit $\mathbf{D}$
and simply write $\mathcal{L}$.
\begin{defn}
SBP is, given a lattice basis $\mathbf{D}_{0}$ of rank $n$, find
\[
\min_{\mathbf{D},\thinspace\mathcal{L}(\mathbf{D})=\mathcal{L}(\mathbf{D}_{0})}\thinspace l(\mathbf{D})
\]
where $l(\mathbf{D})=\max_{i}\left\Vert \mathbf{d}_{i}\right\Vert $,
$\mathbf{D}$ ranges over all possible bases of $\mathcal{L}(\mathbf{D}_{0})$,
and $l(\mathbf{D})$ is referred to as basis length. 
\end{defn}

The Gram-Schmidt orthogonalization (GSO) vectors of a basis $\mathbf{D}$
can be found by: $\mathbf{d}_{1}^{*}=\mathbf{d}_{1},$ $\mathbf{d}_{i}^{*}=\pi_{\mathbf{D}_{\Gamma_{i}}}^{\perp}(\mathbf{d}_{i})=\mathbf{d}_{i}-\sum_{j=1}^{i-1}\mu_{i,j}\mathbf{d}_{j}^{*}$,
for $i=2,\ldots\thinspace,n$, where $\mu_{i,j}=\langle\mathbf{d}_{i},\mathbf{d}_{j}^{*}\rangle/\left\Vert \mathbf{d}_{j}^{*}\right\Vert ^{2}$.
In matrix notations, GSO vectors can be written as $\mathbf{D}=[\mathbf{d}_{1}^{*},\ldots\thinspace,\mathbf{d}_{n}^{*}][\mu_{i,j}]^{\top},$
where $[\mu_{i,j}]$ is a lower-triangular matrix with unit diagonal
elements. In relation to the QR decomposition, let $\Lambda$ be a
diagonal matrix with diagonal entries $\left\Vert \mathbf{d}_{1}^{*}\right\Vert ,\ldots\thinspace,\left\Vert \mathbf{d}_{n}^{*}\right\Vert $,
then we have $[\mathbf{d}_{1}^{*},\ldots\thinspace,\mathbf{d}_{n}^{*}]\Lambda^{-1}=\mathbf{Q}$
and $\Lambda[\mu_{i,j}]^{\top}=\mathbf{R}$ whose diagonal elements
reflect the lengths of GSO vectors. 

The $i$th successive minimum of an $n$ dimensional lattice $\mathcal{L}(\mathbf{D})$
is the smallest real number $r$ such that $\mathcal{L}$ contains
$i$ linearly independent vectors of length at most $r$: 
\[
\lambda_{i}=\inf\left\{ r\mid\dim(\mathrm{span}((\mathcal{L}\cap\mathcal{B}(\mathbf{0},r)))\geq i\right\} ,
\]
in which $\mathcal{B}(\mathbf{t},r)$ denotes a ball centered at $\mathbf{t}$
with radius $r$. We also write $\lambda_{i}$ as $\lambda_{i}(\mathbf{D})$
to distinguish different lattices.

Hermite's constant $\gamma_{n}$ is defined by
\[
\gamma_{n}=\sup_{\mathbf{D}\in\mathbb{R}^{n\times n}}\frac{\lambda_{1}(\mathbf{D})^{2}}{|\mathrm{det}(\mathbf{D})|^{2/n}}.
\]
 Exact values for $\gamma_{n}$ are known for $n\leq8$ and $n=24$.
With Minkowski's convex body theorem, we can obtain $\gamma_{n}\leq\frac{4}{\pi}\Gamma(1+n/2)^{2/n}$,
which yields $\gamma_{n}\leq\frac{2n}{3}$ for $n\geq2$ \cite{Lagarias1990}.
It also follows from the work of Blichfeldt \cite{Blichfeldt1914}
that $\gamma_{n}\leq\frac{2}{\pi}\Gamma(2+n/2)^{2/n}$, whose asymptotic
value is $\frac{n}{\pi e}$. 

The open Voronoi cell of lattice $\mathcal{L}$ with center $\mathbf{v}$
is the set
\[
\mathcal{V}_{\mathbf{v}}(\mathbf{D})=\left\{ \mathbf{x}\in\mathbb{R}^{n}\mid\left\Vert \mathbf{x}-\mathbf{v}\right\Vert <\left\Vert \mathbf{x}-\mathbf{v}-\mathbf{v}'\right\Vert ,\thinspace\forall\mathbf{v}'\in\mathcal{L}\right\} ,
\]
 in which the outer radius of the Voronoi cell centered at the origin
is denoted as ``covering radius'', i.e., $\rho(\mathbf{D})=\max_{\mathbf{t}\in\mathrm{span}(\mathcal{L})}\mathrm{dist}(\mathbf{t},\mathcal{L})$.

The orthogonality defect (OD), $\xi(\mathbf{D})$, can alternatively
quantify the goodness of a basis: 
\begin{equation}
\xi(\mathbf{D})=\frac{\prod_{i=1}^{n}\left\Vert \mathbf{d}_{i}\right\Vert }{\sqrt{\det(\mathbf{D}^{\top}\mathbf{D})}}.\label{eq:OD}
\end{equation}
It has a lower bound $\xi(\mathbf{D})\geq1$ in accordance with Hadamard's
inequality.

\subsection{Lattice reduction algorithms}

In this subsection, we review three popular LR metrics where the lengths
of basis vectors can be upper bounded by scaled versions of the successive
minima. Operations/transforms to reach these metrics are referred
to as the corresponding algorithms. Let $\mathbf{R}$ be the R matrix
of a QR decomposition on $\mathbf{D}$, with elements $r_{i,j}$'s. 
\begin{defn}
A basis $\mathbf{D}$ is called LLL reduced if the following two conditions
hold \cite{Lenstra1982}:

1. $|r_{i,j}/r_{i,i}|\leq\frac{1}{2}$, $1\leq i\leq n$, $j>i$.
(Size reduction conditions) 

2. $\delta\left\Vert \pi_{\mathbf{D}_{\Gamma_{i}}}^{\perp}(\mathbf{d}_{i})\right\Vert ^{2}\leq\left\Vert \pi_{\mathbf{D}_{\Gamma_{i}}}^{\perp}(\mathbf{d}_{i+1})\right\Vert ^{2}$,
$1\leq i\leq n-1$. (Lovász conditions) 
\end{defn}
In the definition, $\delta\in(1/4,1]$ is called the Lovász constant.
If $\mathbf{D}$ is LLL reduced, it has \cite{Lenstra1982} 
\begin{eqnarray}
\left\Vert \mathbf{d}_{i}\right\Vert  & \leq & \beta^{n-1}\lambda_{i}(\mathbf{D}),\thinspace1\leq i\leq n,\label{eq:lll bound}
\end{eqnarray}
 in which $\beta=1/\sqrt{\delta-1/4}\in(2/\sqrt{3},\infty)$.
\begin{defn}
A basis $\mathbf{D}$ is called KZ reduced if it satisfies the size
reduction conditions, and $\pi_{\mathbf{D}_{\Gamma_{i}}}^{\perp}(\mathbf{d}_{i})$
is the shortest vector of the projected lattice $\pi_{\mathbf{D}_{\Gamma_{i}}}^{\perp}([\mathbf{d}_{i},\ldots\thinspace,\mathbf{d}_{n}])$
for $1\leq i\leq n$ \cite{Lagarias1990}. (Projection conditions)
\end{defn}
For a KZ reduced basis, it satisfies \cite{Lagarias1990} 
\begin{equation}
\left\Vert \mathbf{d}_{i}\right\Vert \leq\frac{\sqrt{i+3}}{2}\lambda_{i}(\mathbf{D}),\thinspace1\leq i\leq n.\label{eq:HKZ bound}
\end{equation}

\begin{defn}
A lattice basis $\mathbf{D}$ is called Minkowski reduced if for any
integers $c_{1}$, ..., $c_{n}$ such that $c_{i}$, ..., $c_{n}$
are altogether coprime, it has $\left\Vert \mathbf{d}_{1}c_{1}+\cdots+\mathbf{d}_{n}c_{n}\right\Vert \geq\left\Vert \mathbf{d}_{i}\right\Vert $
for $1\leq i\leq n$ \cite{Zhang2012tsp}. 
\end{defn}
For a Minkowski reduced basis, it satisfies \cite{Zhang2012tsp} 
\begin{equation}
\left\Vert \mathbf{d}_{i}\right\Vert \leq\max\left\{ 1,(5/4)^{(i-4)/2}\right\} \lambda_{i}(\mathbf{D}),\thinspace1\leq i\leq n.\label{eq:min bound}
\end{equation}
When $n\leq4$, Minkowski reduction is optimal as it reaches all the
successive minima. Its bounds on lengths are however exponential for
$n>4$.

\section{Boosted KZ}

In this section, we propose to improve KZ by abandoning its size reduction
conditions, as well as employing the exact closest vector problem
(CVP) oracles to reduce $\mathbf{d}_{i}$ with $\mathcal{L}(\mathbf{D}_{\Gamma_{i}})$
after the projection condition has been met at each time $i$. Better
theoretical results can be obtained via boosted KZ, and the implication
is using CVP for LR can be better than solely relying on SVP. This
should not be a surprise because CVP is generally believed to be harder
than SVP \cite{Micciancio2002}.

\subsection{Replacing size reduction with CVP }

We first show that imposing size reduction conditions in KZ and LLL
may lengthen basis vectors, and thus enlarging OD's. 
\begin{prop}
\label{prop:counterLLL} There always exist real value bases of rank
$n$, $n\geq3$, such that KZ and LLL algorithms lengthen basis vectors.
\footnote{This proposition is inspired by \cite[Lem. 2.2.3]{Nguyen2012b}.}\end{prop}
\begin{IEEEproof}
We prove this by constructing examples in dimension $n=3$ because
bases of higher ranks can be built by concatenating another identity
matrix in the diagonal direction. Consider the following matrix 
\begin{equation}
\mathbf{R}=\left[\begin{array}{ccc}
1 & c_{1} & 0\\
0 & 1 & c_{2}\\
0 & 0 & 1
\end{array}\right],\label{eq:lll fails matrix}
\end{equation}
where $|c_{1}|<1/2$, $|c_{2}|>1/2$. Since $|r_{1,2}/r_{1,1}|<1/2$
and $|r_{2,3}/r_{2,2}|>1/2$, it follows from the definition of KZ
or LLL that $\mathbf{r}_{1}$ and $\mathbf{r}_{2}$ will remain unchanged,
while size reducing $\mathbf{r}_{3}$ by $\mathbf{r}_{2}$ yields
a new vector $\mathbf{r}_{3}'=[-c_{1}\lfloor c_{2}\rceil,c_{2}-\lfloor c_{2}\rceil,1]^{\top}$.
If $\lfloor c_{2}\rceil=\pm1$, then $\mathbf{r}_{3}'$ cannot be
further reduced by $\mathbf{r}_{1}$. So we can assume $\left\Vert \mathbf{r}_{3}'\right\Vert ^{2}>\left\Vert \mathbf{r}_{3}\right\Vert ^{2}$
and solve this inequality about $c_{2}$, which yields $|c_{2}|<\left(1+c_{1}^{2}\right)/2$.
Therefore, there exist at least matrices like (\ref{eq:lll fails matrix})
with $|c_{1}|<1/2$ and $1/2<|c_{2}|<\left(1+c_{1}^{2}\right)/2$
such that KZ/LLL lengthens basis vectors. 
\end{IEEEproof}
To avoid such problems in KZ, we shall review the process of the KZ
reduction algorithm \cite{Zhang2012tsp}. In the beginning, the projection
conditions are met by finding the shortest lattice vectors of the
projected lattices and carrying them to the lattice basis. The size
reduction conditions are subsequently addressed by using Babai points
$\mathbf{v}$'s in $\mathcal{L}(\mathbf{D}_{\Gamma_{i}})$ to reduce
$\mathbf{d}_{i}$ by $\mathbf{d}_{i}\leftarrow\mathbf{d}_{i}-\mathbf{v}$
for all $i$. Concerning the above procedure, what we try to ameliorate
are the size reduction operations. The ``$\mathbf{d}_{i}\leftarrow\mathbf{d}_{i}-\mathbf{v}$''
step is redefined as \textit{length reduction}, in which the optimal
update needs to solve a CVP.
\begin{defn}
CVP is a problem that, given a vector $\mathbf{y}\in\mathbb{R}^{n}$
and a lattice basis $\mathbf{D}$ of rank $n$, find a vector $\mathbf{v}\in\mathcal{L}(\mathbf{D})$
such that $\left\Vert \mathbf{y}-\mathbf{v}\right\Vert ^{2}\leq\left\Vert \mathbf{y}-\mathbf{w}\right\Vert ^{2},\thinspace\forall\thinspace\mathbf{w}\in\mathcal{L}(\mathbf{D})$. 
\end{defn}
An algorithm solving CVP, which quantizes any input to a lattice point,
is denoted as $\mathbf{v}=\mathcal{Q}_{\mathcal{L}(\mathbf{D})}(\mathbf{y})$.
It is evident that $\mathcal{Q}_{\mathcal{L}(\mathbf{D}_{\Gamma_{i}})}(\pi_{\mathbf{D}_{\Gamma_{i}}}(\mathbf{d}_{i}))=\mathcal{Q}_{\mathcal{L}(\mathbf{D}_{\Gamma_{i}})}(\mathbf{d}_{i})$
. To obtain explicit properties from the length reductions, we first
establish Proposition \ref{thm:Voroni small} to show
\[
\left\Vert \mathbf{d}_{i}-\mathcal{Q}_{\mathcal{L}(\mathbf{D}_{\Gamma_{i}})}(\pi_{\mathbf{D}_{\Gamma_{i}}}(\mathbf{d}_{i}))\right\Vert <\left\Vert \mathbf{d}_{i}\right\Vert 
\]
if $\mathcal{Q}_{\mathcal{L}(\mathbf{D}_{\Gamma_{i}})}(\pi_{\mathbf{D}_{\Gamma_{i}}}(\mathbf{d}_{i}))\neq\mathbf{0}$
for all $i$. The proof is given in Appendix \ref{sec: short quan}. 
\begin{prop}
\label{thm:Voroni small} If $\pi_{\mathbf{D}_{\Gamma_{i}}}(\mathbf{d}_{i})$
lies outside the Voronoi region $\mathcal{V}_{\mathbf{0}}(\mathbf{D}_{\Gamma_{i}})$,
i.e., $\mathbf{v}\triangleq\mathcal{Q}_{\mathcal{L}(\mathbf{D}_{\Gamma_{i}})}(\pi_{\mathbf{D}_{\Gamma_{i}}}(\mathbf{d}_{i}))\neq\mathbf{0}$,
then we can replace $\mathbf{d}_{i}$ with $\mathbf{d}_{i}-\mathbf{v}$
because $\left\Vert \mathbf{d}_{i}-\mathbf{v}\right\Vert <\left\Vert \mathbf{d}_{i}\right\Vert $.
\end{prop}
Together with the case of $\mathcal{Q}_{\mathcal{L}(\mathbf{D}_{\Gamma_{i}})}(\pi_{\mathbf{D}_{\Gamma_{i}}}(\mathbf{d}_{i}))=\mathbf{0}$,
we conclude that 
\begin{equation}
\left\Vert \mathbf{d}_{i}-\mathcal{Q}_{\mathcal{L}(\mathbf{D}_{\Gamma_{i}})}(\pi_{\mathbf{D}_{\Gamma_{i}}}(\mathbf{d}_{i}))\right\Vert \leq\left\Vert \mathbf{d}_{i}\right\Vert \label{eq:reduced svp}
\end{equation}
for all $i$, which means, during the length reductions, all solutions
provided by CVP can be treated as effective updates. We call them
effective because each $\mathbf{d}_{i}$ is the shortest vector that
can be extended to a basis for $\mathcal{L}([\mathbf{D}_{\Gamma_{i}},\mathbf{d}_{i}])$,
and the length reductions never increase the lengths of $\mathbf{d}_{i}$'s.

After executing these length reduction operations as $\mathbf{d}_{i}\leftarrow\mathbf{d}_{i}-\mathcal{Q}_{\mathcal{L}(\mathbf{D}_{\Gamma_{i}})}(\pi_{\mathbf{D}_{\Gamma_{i}}}(\mathbf{d}_{i}))$,
all $\pi_{\mathbf{D}_{\Gamma_{i}}}(\mathbf{d}_{i})$'s must lie inside
the Voronoi regions $\mathcal{V}_{\mathbf{0}}(\mathbf{D}_{\Gamma_{i}})$'s,
so that 
\begin{equation}
\left\Vert \mathbf{d}_{i}\right\Vert ^{2}\leq\left\Vert \pi_{\mathbf{D}_{\Gamma_{i}}}^{\perp}(\mathbf{d}_{i})\right\Vert ^{2}+\rho(\mathbf{D}_{\Gamma_{i}})^{2}\label{eq:common equation}
\end{equation}
for all $i$, where $\rho(\mathbf{D}_{\Gamma_{i}})$ is the covering
radius of $\mathcal{L}(\mathbf{D}_{\Gamma_{i}})$.

\subsection{Algorithm description}

The concrete steps of boosted KZ are presented in Algorithm \ref{algoKZ}.
This algorithm can be briefly explained as follows. In line 4, the
Schnorr and Euchner (SE) enumeration algorithm \cite{Schnorr1994}
is applied to solve SVP over $\mathcal{L}(\mathbf{R}_{i:n,i:n})$,
in that if $\mathbf{R}_{i:n,i}$ is the shortest vector of $\mathcal{L}(\mathbf{R}_{i:n,i:n})$,
then $\pi_{\mathbf{D}_{\Gamma_{i}}}^{\perp}(\mathbf{d}_{i})$ is the
shortest vector of the projected lattice $\pi_{\mathbf{D}_{\Gamma_{i}}}^{\perp}([\mathbf{d}_{i},\ldots\thinspace,\mathbf{d}_{n}])$.
Lines 5 to 7 are designed to plug new vectors found into the lattice
basis, and the basis expansion method in \cite{Zhang2012tsp} can
do this efficiently. Other basis expansion methods include using LLL
reduction \cite{Chen2011b} or employing the Hermite normal form of
the coefficient matrix \cite[Lem. 7.1]{Micciancio2002}, but both
of them have higher complexity than the one in \cite{Zhang2012tsp}.
Lines 8 to 10 restore the upper triangular property of $\mathbf{R}$,
and these be alternatively implemented by performing another QR decomposition.
\textit{Line 11 is the unique new design of boosted KZ}, i.e., to
reduce $\mathbf{R}_{1:n,i}$ by using its closest vector in $\mathcal{L}(\mathbf{R}_{1:n,1:i-1})$.\footnote{Since we only modify the size reduction steps in KZ, one may employ
any improved KZ implementation, e.g., \cite{Wen2015}, to make the
boosted KZ faster. We adhere to the current version for making a fair
complexity comparison with Minkowski's reduction which employs similar
subroutines \cite{Zhang2012tsp}.}

\begin{algorithm} \caption{The boosted KZ algorithm.}
\label{algoKZ}
\KwIn{original lattice basis $\mathbf{D}\in \mathbb{R}^{n\times n}$, Lovász constant $\delta$.} \KwOut{reduced basis $\mathbf{D}$, unimodular matrix $\mathbf{T}$}   $[\mathbf{Q},\mathbf{R}]=\mathrm{qr}(\mathbf{D})$;\Comment{The QR decomposition of $\mathbf{D}$}\;  
 $\mathbf{T}=\mathbf{I}$\; 
\For{$i=1:n$}     {   
find the shortest  vector $\mathbf{R}_{i:n, i:n}\mathbf{c}_1$ in $\mathcal{L}(\mathbf{R}_{i:n, i:n})$ by  LLL  aided  SE enumeration;\Comment{SVP subroutine}\; 
construct a $(n-i+1)\times (n-i+1)$ unimodular matrix $\mathbf{U}$ whose first column is  $\mathbf{c}_1$\; $\mathbf{R}_{1:n, i:n} \leftarrow \mathbf{R}_{1:n, i:n}\mathbf{U}$\; 
$\mathbf{T}_{1:n, i:n}\leftarrow \mathbf{T}_{1:n, i:n}\mathbf{U}$\; define $\mathbf{G}$ as a unitary matrix that can restore the upper triangular  property of $\mathbf{R}$\; $\mathbf{R}\leftarrow \mathbf{G}\mathbf{R}$\; $\mathbf{Q}\leftarrow \mathbf{Q}\mathbf{G}^\top$\; 
find the closest  vector $\mathbf{R}_{1:n, 1:i-1}\mathbf{c}_2$ in $\mathcal{L}(\mathbf{R}_{1:n, 1:i-1})$ to $\mathbf{R}_{1:n,i}$ with SE enumeration; \Comment{CVP subroutine}\;
$\mathbf{R}_{1:n, i}\leftarrow \mathbf{R}_{1:n, i}-\mathbf{R}_{1:n, 1:i-1}\mathbf{c}_2$\; 
$\mathbf{T}_{1:n, i}\leftarrow \mathbf{T}_{1:n, i}-\mathbf{T}_{1:n, 1:i-1}\mathbf{c}_2$\;
} $\mathbf{D}\leftarrow \mathbf{Q}\mathbf{R}.$ \end{algorithm}

\subsection{Properties of boosted KZ}

Based on Algorithm \ref{algoKZ}, a lattice basis $\mathbf{D}$ is
called boosted KZ reduced if $\pi_{\mathbf{D}_{\Gamma_{i}}}^{\perp}(\mathbf{d}_{i})$
is the shortest vector of the projected lattice $\pi_{\mathbf{D}_{\Gamma_{i}}}^{\perp}([\mathbf{d}_{i},\ldots\thinspace,\mathbf{d}_{n}])$,
and $\pi_{\mathbf{D}_{\Gamma_{i}}}(\mathbf{d}_{i})\in\mathcal{V}_{\mathbf{0}}(\mathbf{D}_{\Gamma_{i}})$
for all $i$. 

In boosted KZ, all length reductions are the strongest, and they can
help us to deliver better bounds for the lengths of basis vectors,
as given in Proposition \ref{prop:KZ len}. The proof is given in
Appendix \ref{sec: kz len}. We have $\left\Vert \mathbf{d}_{n}\right\Vert \leq\max\left\{ 1,\thinspace\frac{\sqrt{n}}{2}\right\} \lambda_{n}(\mathbf{D})$,
outperforming the $\left\Vert \mathbf{d}_{n}\right\Vert \leq\frac{\sqrt{n+3}}{2}\lambda_{n}(\mathbf{D})$
bound in \cite[Thm. 2.1]{Lagarias1990} which was conjectured not
tight in their work. 
\begin{prop}
\label{prop:KZ len}Suppose a basis $\mathbf{D}$ is boosted KZ reduced,
then this basis satisfies
\begin{equation}
\left\Vert \mathbf{d}_{i}\right\Vert \leq\min\left\{ \frac{\sqrt{i+3}}{2}\lambda_{i}(\mathbf{D}),\max\left\{ 1,\thinspace\frac{\sqrt{i}}{2}\right\} \lambda_{i}(\mathbf{D}_{\Gamma_{i+1}})\right\} \label{eq:len1}
\end{equation}
for $1\leq i<n$, and 
\begin{equation}
\left\Vert \mathbf{d}_{n}\right\Vert \leq\max\left\{ 1,\thinspace\frac{\sqrt{n}}{2}\right\} \lambda_{n}(\mathbf{D}).\label{eq:len2-1}
\end{equation}

\end{prop}
A direct application of the above proposition also shows a boosted
KZ reduced basis has length 
\begin{equation}
l(\mathbf{D})\leq\frac{\sqrt{n+2}}{2}\lambda_{n}(\mathbf{D}).\label{eq:l hkz}
\end{equation}

\begin{rem}
Our results of (\ref{eq:len1}), (\ref{eq:len2-1}) and (\ref{eq:l hkz})
are better than those of KZ and Minkowski reductions. If we assume
all the successive minima are available, then there exists a polynomial
time transformation that generates a basis with $l(\mathbf{D})\leq\max\left\{ 1,\sqrt{n}/2\right\} \lambda_{n}(\mathbf{D})$
\cite[Lem. 7.1]{Micciancio2002}. 
\end{rem}
The lengths of GSO vectors in this new algorithm remains the same
as those of KZ reduction, so readily we can claim that $\lambda_{1}(\mathbf{D})^{2}\leq i^{1+\ln(i)}r_{i,i}^{2}$
and $\left\Vert \mathbf{d}_{i}\right\Vert ^{2}\leq i^{2+\ln(i)}r_{i,i}^{2}$
as given in \cite[Prop. 4.2]{Lagarias1990}, where $r_{i,i}$ denotes
the $(i,\thinspace i)$th entry of the $\mathbf{R}$ matrix of a QR
decomposition on $\mathbf{D}$. As another contribution, now we show
these two bounds can be improved in Proposition \ref{prop:gs lambda1 },
whose proof is given in Appendix \ref{sec: kz gsgap}.
\begin{prop}
\label{prop:gs lambda1 }Suppose a basis $\mathbf{D}$ is boosted
KZ reduced, then this basis satisfies

\begin{equation}
\lambda_{1}(\mathbf{D})^{2}\leq\frac{8i}{9}(i-1)^{\ln(i-1)/2}r_{i,i}^{2},\label{eq:lambda1 gslen}
\end{equation}

\textup{
\begin{equation}
\left\Vert \mathbf{d}_{i}\right\Vert ^{2}\leq\Big(1+\frac{2i}{9}(i-1)^{1+\ln(i-1)/2}\Big)r_{i,i}^{2}\label{eq:len gs3}
\end{equation}
}for $1\leq i\leq n$.
\end{prop}
The relaxed versions of (\ref{eq:lambda1 gslen}) and (\ref{eq:len gs3})
can be read as $\lambda_{1}(\mathbf{D})^{2}\leq i^{1+\ln(i)/2}r_{i,i}^{2}$
and $\left\Vert \mathbf{d}_{i}\right\Vert ^{2}\leq i^{2+\ln(i)/2}r_{i,i}^{2}$.
Proposition \ref{prop:gs lambda1 } can be either applied to bound
the complexity of boosted KZ, or to achieve the best explicit bounds
for the proximity factors of lattice reduction aided decoding, i.e.,
updating Eqs. (41) and (45) of \cite{Ling2011}. Moreover, (\ref{eq:len gs3})
leads to an alternative bound for OD,
\[
\xi(\mathbf{D})\leq\prod_{i=1}^{n}i^{1+\ln(i)/4}\leq n^{n+\ln(n!)/4}.
\]
 A better bound on $\xi(\mathbf{D})$ comes after applying Minkowski's
second theorem \cite[P. 202]{Cassels1971} to (\ref{eq:len1}) and
(\ref{eq:len2-1}),

\begin{equation}
\xi(\mathbf{D})\leq\frac{\sqrt{n}}{2}\left(\prod_{i=1}^{n-1}\frac{\sqrt{i+3}}{2}\right)\left(\frac{2}{3}n\right)^{n/2}.\label{eq:od bHKZ}
\end{equation}

\begin{rem}
The properties of $|r_{i,j}/r_{i,i}|\leq\frac{1}{2}$ for $1\leq i\leq n$,
$j>i$, are no longer guaranteed in boosted KZ. Of independent interests,
we have another attribute in Proposition \ref{claim: PR properties}
that\textbf{ }each pair $(\mathbf{d}_{1},\mathbf{d}_{i})$ of the
boosted KZ reduced basis is Lagrange reduced \cite[P. 41]{Nguyen2010}
for all $i$, which may not hold in the conventional KZ. The proof
can be found in Appendix \ref{sec: gauss red}.\end{rem}
\begin{prop}
\label{claim: PR properties} Suppose a basis $\mathbf{D}$ is boosted
KZ reduced, then this basis satisfies $|r_{1,i}/r_{1,1}|\leq\frac{1}{2}$,
and $\left\Vert \mathbf{d}_{1}\right\Vert $, $\left\Vert \mathbf{d}_{i}\right\Vert $
reaches the first and second successive minima of $\mathcal{L}([\mathbf{d}_{1},\mathbf{d}_{i}])$
for $2\leq i\leq n$.
\end{prop}

\subsection{Implementation and complexity}

The complexity of boosted KZ is dominated by its SVP and CVP subroutines,
in which the SE enumeration algorithm \cite{Schnorr1994} will be
adopted for our implementations and complexity analysis. The total
complexity is assessed by counting the number of floating-point operations
(flops).

\subsubsection{Complexity of CVP subroutines}

It suffices to discuss the complexity of the most time-consuming $n$th
round of reducing $\mathbf{R}_{1:n,n}$, which represents an $n-1$
dimensional CVP problem. First of all, the complexity of SE is directly
proportional to the number of nodes in the search tree. In the $k$th
layer of the enumeration, the number of nodes is 
\begin{multline*}
N_{k}(s)=\Big\{|\mathbf{x}_{k:n-1}|\mid\mathbf{x}_{k:n-1}\in\mathbb{Z}^{n-k},\\
\left\Vert \mathbf{R}_{k:n-1,n}-\mathbf{R}_{k:n-1,k:n-1}\mathbf{x}_{k:n-1}\right\Vert ^{2}\leq s^{2}\Big\}
\end{multline*}
where $s$ refers to the radius of a specified sphere and $\mathbf{R}_{1:n-1,n}$
is the projection of $\mathbf{R}_{1:n,n}$ onto $\mathrm{span}(\mathbf{R}_{1:n-1,1:n-1})$.
From \cite{Chang2013}, $N_{k}(s)$ can be estimated by $N_{k}(s)\approx\frac{V_{n-k}(1)s^{n-k}}{|r_{k,k}|\cdots|r_{n-1,n-1}|}$
where $V_{n}(1)=\frac{\pi^{n/2}}{\Gamma(1+n/2)}\sim\left(\frac{2e\pi}{n}\right)^{n/2}\frac{1}{\sqrt{\pi n}}$
stands for the volume of an $n$ dimensional unit ball. Since $\lim_{n\rightarrow\infty}V_{n}(1)=0$,
we can cancel this term in the asymptotic analysis. By summing the
nodes from layer $1$ to $n-1$, the total number of nodes in the
$n-1$ dimensional CVP can be given as 
\[
N_{\mathrm{CVP},n-1}(s)=\sum_{k=1}^{n-1}\frac{V_{n-k}(1)s^{n-k}}{|r_{k,k}|\cdots|r_{n-1,n-1}|}.
\]

For the visit of each node, the operations of updating the residual
and outer radius etc. cost around $2k+7$ flops in layer $k$, so
the complexity of the $n-1$ dimensional CVP problem, $F_{\mathrm{CVP},n-1}(s)$,
can be accessed:

\begin{multline}
F_{\mathrm{CVP},n-1}(s)=\sum_{k=1}^{n-1}N_{k}(s)(2k+7)\\
\leq\sum_{k=1}^{n-1}\frac{V_{n-k}(1)s^{n-k}\prod_{j=k}^{n-1}\thinspace j^{1/2+\ln(j)/4}(2k+7)}{\lambda_{1}(\mathbf{D})^{n-k}},\label{eq:number CVP}
\end{multline}
 in which Proposition \ref{prop:gs lambda1 } has been used to get
the inequality. 

Then we present a general strategy to choose $s$. It starts from
$s=|r_{1,1}|/2$ which equals to the packing radius because $\lambda_{1}(\mathbf{R}_{1:n,1:k-1})=|r_{1,1}|$,
and improves $s$ to $\frac{1}{2}\sqrt{\sum_{j=1}^{k}r_{j,j}^{2}}$
with $k=2,\thinspace\ldots,\thinspace n-1$ gradually until at least
one node can be found inside the searching sphere. For a random basis,
one may expect $s=|r_{1,1}|/2$ to work well with high probability,
though we can also use the worst case criterion of $s=\frac{1}{2}\sqrt{\sum_{j=1}^{n-1}r_{j,j}^{2}}$
(i.e., larger than the covering radius). In the worst case, let $C_{1}=\frac{\sqrt{n-1}\lambda_{n}(\mathbf{D})}{2\lambda_{1}(\mathbf{D})}$.
Since $V_{n}(1)(2n+7)$ also vanishes for large $n$, then (\ref{eq:number CVP})
can be written as
\begin{equation}
F_{\mathrm{CVP},n-1}(C_{1}\lambda_{1}(\mathbf{D}))\leq nC_{1}^{n}n^{n/2+\ln(n!)/4}\label{eq:number CVP2}
\end{equation}

\subsubsection{Complexity of SVP subroutines}

Among the SVP subroutines of boosted KZ, its first round of finding
$\lambda_{1}(\mathbf{D})$ is the most difficult one. By invoking
the Siegel condition of $|r_{i-1,i-1}|\leq\beta|r_{i,i}|$ due to
LLL \cite{Gama2006}, we have $1/\prod_{j=k}^{n}|r_{j,j}|\leq\beta^{(n+k-2)(n-k+1)/2}/\lambda_{1}(\mathbf{D})^{n-k+1}$,
so the number of flops spent by the first round SVP subroutine can
be similarly bounded as 
\begin{multline}
F_{\mathrm{SVP},n}(s)\leq\\
\sum_{k=1}^{n}\frac{V_{n-k+1}(1)s^{n-k+1}\beta^{(n+k-2)(n-k+1)/2}(2k+7)}{\lambda_{1}(\mathbf{D})^{n-k+1}}.\label{eq:first LLLHKZ}
\end{multline}
A practical principle for choosing $s$ is to set $s=\left\Vert \mathbf{R}_{1:n,1}\right\Vert $,
which is no smaller than $\lambda_{1}(\mathbf{D})$. It follows from
an LLL reduced basis property of $\left\Vert \mathbf{R}_{1:n,1}\right\Vert \leq\beta^{n-1}\lambda_{1}(\mathbf{D})$
that (\ref{eq:first LLLHKZ}) becomes 
\begin{equation}
F_{\mathrm{SVP},n}(\beta^{n-1}\lambda_{1}(\mathbf{D}))\leq n\beta^{n(n-1)}\beta^{n(n-1)/2}=n\beta^{3n(n-1)/2}.\label{eq:svp sub}
\end{equation}

\subsubsection{Total complexity in flops}

The complexity of other operations in Algorithm \ref{algoKZ} can
be counted as well. In the $i$th round, Lines 5 to 10 being implemented
by the method described in \cite[Fig. 3]{Zhang2012tsp} costs $O(n(n-i))$.
The total complexity of boosted KZ is therefore upper bounded as

\begin{multline}
F_{\mathrm{boostKZ}}\leq(n-1)\Big(F_{\mathrm{CVP},n-1}(C_{1}\lambda_{1}(\mathbf{D}))+F_{\mathrm{SVP},n}(\beta^{n-1}\lambda_{1}(\mathbf{D}))\\
+O(n^{2}-n)\Big)+\frac{4}{3}n^{3}+(2n-1)n^{2}.\label{eq:lasttotal}
\end{multline}
By plugging (\ref{eq:number CVP2}) and (\ref{eq:svp sub}) inside
(\ref{eq:lasttotal}), we can explicitly obtain
\[
F_{\mathrm{boostKZ}}\leq C_{1}^{n}n^{n/2+\ln(n!)/4+O(\ln n)^{2}}+\beta^{3n(n-1)/2+O(\ln n)^{2}}.
\]

A few comments are made regarding the above analysis. Firstly, it
provides a worst case analysis for strong lattice reductions like
KZ and boosted KZ, which is broad enough to include many applications.
A byproduct of our analysis is that we can replace the term $F_{\mathrm{CVP},n-1}(C_{1}\lambda_{1}(\mathbf{D}))$
in (\ref{eq:lasttotal}) with $O(n^{3})$ to get the worst case complexity
of KZ, i.e., $F_{\mathrm{KZ}}\leq\beta^{3n(n-1)/2+O(\ln n)^{2}}$.
This compensates the expected complexity analysis in \cite[Sec. III.C]{Zhang2012tsp}
which hinges on Gaussian lattice bases. Secondly, we can observe from
(\ref{eq:lasttotal}) that how much harder the boosted KZ has become
by using CVP. If $\lambda_{1}(\mathbf{D})$ is of the same order as
$\lambda_{n}(\mathbf{D})$, we can put $C_{1}\approx\frac{\sqrt{n-1}}{2}$
into (\ref{eq:lasttotal}) to conclude that boosted KZ is not much
more complicated than KZ. Actually, if the lattices are random (see
\cite[Sec. 2]{Nguyen2006} for more details about random lattices),
then the Gaussian heuristic implies $\lambda_{1}(\mathbf{D})\approx\cdots\approx\lambda_{n}(\mathbf{D})$
\cite{Gama2008}. In the application to IF, our claim that boosted
KZ is not much harder than KZ will be supported by simulations in
Section V.

\section{Boosted LLL}

In the same spirit of extending size reductions to length reductions,
we will revamp LLL towards better performance in this section. In
a nutshell, the boosted LLL algorithm implements its length reduction
via the parallel nearest plane (PNP) algorithm \cite[Sec. 4]{Lindner2011}
and rejection. PNP can be regarded as a compromise between Babai's
nearest plane algorithm and the CVP oracle. If PNP has a route number
$L=1$, then it becomes equivalent to Babai's algorithm, while setting
$L$ infinitely large solves CVP. The complexity of boosted LLL is
about $L$ times as large as the that of LLL. In our algorithm, setting
$L=1$ means only imposing a rejection operation.

\subsection{Replacing size reduction with PNP and rejection}

\begin{figure}[tbh]
\center

\includegraphics[width=3.4in,height=2.6in]{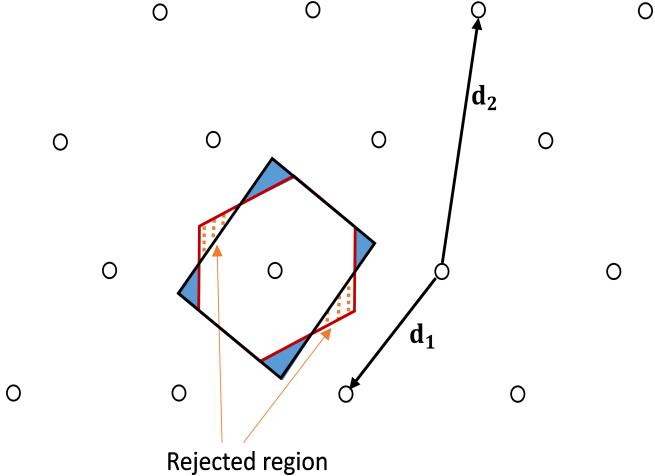}

\protect\caption{The rejected region (orange dots) of the possible projections $\pi_{\mathbf{D}_{\Gamma_{3}}}(\mathbf{d}_{3})$
in $\mathcal{L}(\mathbf{D}_{\Gamma_{3}})$ with respect to $\mathcal{V}_{\mathbf{0}}(\mathbf{D}_{\Gamma_{3}})$
(red hexagon) and $\mathcal{P}(\mathbf{d}_{1}^{*},\mathbf{d}_{2}^{*})$
(black rectangle), where the size reduction of LLL elongates \textbf{$\mathbf{d}_{3}$}.
The four blue triangles are the region whose Babai point is the origin
and size reductions cannot alter a suboptimal $\mathbf{d}_{3}$.}

\label{figReject}
\end{figure}

First of all, the classic LLL algorithm consists of two sequential
phases, i.e., size reductions by using Babai points, and swaps based
on testing the Lovász conditions. To reduce $\mathbf{d}_{i}$ with
$\mathbf{D}_{\Gamma_{i}}$, the sharpest reduction should utilize
the closest vector of $\pi_{\mathbf{D}_{\Gamma_{i}}}(\mathbf{d}_{i})$
in $\mathcal{L}(\mathbf{D}_{\Gamma_{i}})$ as shown in Proposition
\ref{thm:Voroni small}. In order to devote flexible efforts to these
length reductions, we shall investigate the success probability of
a Babai point being optimum. Generally, assume $\pi_{\mathbf{D}_{\Gamma_{i}}}(\mathbf{d}_{i})$
is uniformly distributed over $\mathcal{L}(\mathbf{D}_{\Gamma_{i}})$,
then the probability of a Babai point being the closest vector is 

\begin{equation}
\frac{\int_{\mathbf{x}\in\mathcal{V}_{\mathbf{0}}(\mathbf{D}_{\Gamma_{i}})}\mathbb{I}(\mathbf{x}\in\mathcal{P}(\mathbf{D}_{\Gamma_{i}}^{*}))\mathrm{d}\mathbf{x}}{|\mathcal{V}_{\mathbf{0}}(\mathbf{D}_{\Gamma_{i}})|}=\frac{|\mathcal{V}_{\mathbf{0}}(\mathbf{D}_{\Gamma_{i}})\cap\mathcal{P}(\mathbf{D}_{\Gamma_{i}}^{*})|}{|\mathcal{V}_{\mathbf{0}}(\mathbf{D}_{\Gamma_{i}})|},\label{eq:mul sic}
\end{equation}
 in which $\mathcal{P}(\mathbf{D}_{\Gamma_{i}}^{*})=\left\{ \sum_{k=1}^{i-1}c_{k}\mathbf{d}_{k}^{*}\thinspace|\thinspace-1/2\leq c_{k}\leq1/2\right\} $
is the parallelepiped of GSO vectors $\left\{ \mathbf{d}_{1}^{*},\ldots\thinspace,\mathbf{d}_{i-1}^{*}\right\} $,
and $\mathbb{I}(\cdot)$ denotes an indicator function. One evident
observation from Eq. (\ref{eq:mul sic}) is, by updating $\mathbf{d}_{1}^{*}\leftarrow p_{1}\mathbf{d}_{1}^{*},\thinspace\ldots\thinspace,\mathbf{d}_{i-1}^{*}\leftarrow p_{i-1}\mathbf{d}_{i-1}^{*},$
the probability in Eq. (\ref{eq:mul sic}) rises if we choose some
constants $p_{1}>1,\ldots\thinspace,p_{i-1}>1$. Another implication
from Eq. (\ref{eq:mul sic}) is, if $\pi_{\mathbf{D}_{\Gamma_{i}}}(\mathbf{d}_{i})$
belongs to both the external of $\mathcal{P}(\mathbf{D}_{\Gamma_{i}}^{*})$
and internal of $\mathcal{V}_{\mathbf{0}}(\mathbf{D}_{\Gamma_{i}})$,
then a Babai point should be rejected; otherwise it elongates $\mathbf{d}_{i}$.
An example of $i=3$ is shown in Fig. \ref{figReject}. 

With the above demonstrations, we propose to amplify the success probabilities
of Babai points with minimal efforts and to reject operations that
elongate current basis vectors. Either using lattice Gaussian sampling
\cite{Liu2011it} or PNP suffices the first objective, but we shall
adhere to PNP because it is deterministic and this feature will be
employed by (\ref{eq:condition1-1}). We detail the length reductions
in boosted LLL as follows. 

Assume we are working on the $\mathbf{R}$ matrix of a QR decomposition
and trying to reduce $\mathbf{r}_{i}$ by $\mathbf{r}_{i-1},\ldots\thinspace,\mathbf{r}_{1}$.
Let PNP be abstracted by a parameter $L=\prod_{k=1}^{i-1}p_{k}$ indicating
the total number of routes it consists of, and $(p_{i-1},\ldots\thinspace,p_{1})\in(\mathbb{Z}^{+})^{i-1}$.
Then each route of PNP can be marked by a label $(q_{i-1},\ldots\thinspace,q_{1})$
where $(q_{i-1},\ldots\thinspace,q_{1})\in\left\{ 1,\ldots\thinspace,p_{i-1}\right\} \times\cdots\times\left\{ 1,\ldots\thinspace,p_{1}\right\} $.
From layer $i-1$ of each route, let $c_{i-1,(q_{i-1},...,q_{1})}$
be the $q_{i-1}$th closest integer to $\lfloor r_{i-1,i}/r_{i-1,i-1}\rceil$,
and $\mathbf{r}_{i,(q_{i-1},...,q_{1})}=\mathbf{r}_{i}$, we set $\mathbf{r}_{i,(q_{i-1},...,q_{1})}\leftarrow\mathbf{r}_{i,(q_{i-1},...,q_{1})}-c_{i-1,(q_{i-1},...,q_{1})}\mathbf{r}_{i-1}$
and repeat this process down to layer $1$, resulting in $L$ pairs
of coefficient vectors $\mathbf{c}_{(q_{i-1},...,q_{1})}=[-c_{1,(q_{i-1},...,q_{1})},\ldots\thinspace,-c_{i-1,(q_{i-1},...,q_{1})},1]$
and residuals $\mathbf{r}_{i,(q_{i-1},...,q_{1})}=\mathbf{R}_{\Gamma_{i+1}}\mathbf{c}_{(q_{i-1},...,q_{1})}$.
We also mark the old $\mathbf{r}_{i}$ by $\mathbf{r}_{i,(0,\cdots,0)}=\mathbf{r}_{i}$,
and $\mathbf{c}_{(0,\cdots,0)}=[0,\ldots\thinspace,0,1]$. At this
stage, it can choose the shortest vector among all the $L+1$ candidates
as the reduced version of $\mathbf{r}_{i}$, i.e., $\mathbf{r}_{i}=\mathbf{r}_{i,(z_{i-1}^{*},...,z_{1}^{*})}$
where 
\begin{equation}
(z_{i-1}^{*},...,z_{1}^{*})=\arg\min_{(z_{i-1},...,z_{1})}\left\{ \left\Vert \mathbf{r}_{i,(z_{i-1},...,z_{1})}\right\Vert \right\} .\label{eq:condition1}
\end{equation}
 If one also intends to export the unimodular transformation matrix
$\mathbf{T}$, then it can be simultaneously updated inside PNP, which
means $\mathbf{t}_{i,(z_{i-1}^{*},...,z_{1}^{*})}=\mathbf{T}_{\Gamma_{i+1}}\mathbf{c}_{(z_{i-1}^{*},...,z_{1}^{*})}$,
$\mathbf{T}_{1:n,1:i}=\mathbf{t}_{i,(z_{i-1}^{*},...,z_{1}^{*})}$.

Since $|r_{k,i}/r_{k,k}|<1/2$ for $k<i$ is no longer guaranteed,
together with the Lovász condition they may destroy the Siegel condition
\cite{Gama2006} $r_{i-1,i-1}^{2}\leq(\frac{4}{3}+\varepsilon)r_{i,i}^{2}$
for $2\leq i\leq n$ with some small $\varepsilon>0$. For this reason,
we should relax the Lovász condition to the diagonal reduction (DR)
condition \cite{Zhang2012}.
\begin{defn}[DR condition \cite{Zhang2012}]
 An upper triangular lattice basis $\mathbf{R}$ satisfies the DR
condition with parameter $\delta\thinspace(1/2<\delta<1)$ if it has
\begin{equation}
\delta r_{i-1,i-1}^{2}\leq r_{i,i}^{2}+(r_{i-1,i}-\lfloor r_{i-1,i}/r_{i-1,i-1}\rceil r_{i-1,i-1})^{2}\label{eq:diagonal condition}
\end{equation}
 for all $2\leq i\leq n$, where $\delta$ is still referred to as
the Lovász constant.
\end{defn}
If (\ref{eq:diagonal condition}) holds, the Siegel condition must
be true, so we let $i\leftarrow i+1$ and safely go to the next iteration.
However, if (\ref{eq:diagonal condition}) fails, one should also
investigate whether a swap can tweak such cases. Consider the sublattice
$\mathcal{L}(\mathbf{R}_{\Gamma_{i+1}})$ generated by the first $i$
vectors and define the potential of basis $\mathbf{R}$ \cite{Lenstra1982}
as 
\begin{equation}
\mathrm{Pot}(\mathbf{R})=\prod_{i=1}^{n}\mathrm{det}(\mathcal{L}(\mathbf{R}_{\Gamma_{i+1}}))^{2}=\prod_{i=1}^{n}r_{i,i}^{2(n-i+1)}.\label{eq:def pot}
\end{equation}
If the DR condition fails in $\pi_{\mathbf{R}_{\Gamma_{i}}}^{\perp}(\mathbf{R}_{1:n,i-1:i})$
and we swap $\mathbf{r}_{i-1}$ and $\mathbf{r}_{i}$, then the potential
of the basis should be decreasing for the sake of bounding the number
of iterations. After the swap, $\mathbf{R}_{i-1:i,i-1:i}$ becomes
\begin{equation}
\left[\begin{array}{cc}
r_{i-1,i} & r_{i-1,i-1}\\
r_{i,i} & 0
\end{array}\right].\label{eq:upper1}
\end{equation}
 Let $\mathbf{G}$ be a $2\times2$ unitary matrix 
\begin{equation}
\left[\begin{array}{cc}
\frac{r_{i-1,i}}{\sqrt{r_{i,i}^{2}+r_{i-1,i}^{2}}} & \frac{r_{i,i}}{\sqrt{r_{i,i}^{2}+r_{i-1,i}^{2}}}\\
-\frac{r_{i,i}}{\sqrt{r_{i,i}^{2}+r_{i-1,i}^{2}}} & \frac{r_{i-1,i}}{\sqrt{r_{i,i}^{2}+r_{i-1,i}^{2}}}
\end{array}\right];\label{eq:given matrix}
\end{equation}
clearly, $\mathbf{G}\mathbf{R}_{i-1:i,1:n}$ can restore the upper
triangular property of (\ref{eq:upper1}), which transforms to
\begin{equation}
\left[\begin{array}{cc}
\sqrt{r_{i,i}^{2}+r_{i-1,i}^{2}} & \frac{r_{i-1,i}r_{i-1,i-1}}{\sqrt{r_{i,i}^{2}+r_{i-1,i}^{2}}}\\
0 & -\frac{r_{i,i}r_{i-1,i-1}}{\sqrt{r_{i,i}^{2}+r_{i-1,i}^{2}}}
\end{array}\right].\label{eq:new block}
\end{equation}
From (\ref{eq:def pot}) and (\ref{eq:new block}), one can obtain
the potential ratio between two consecutive bases $\mathbf{R}'$ and
$\mathbf{R}$ as
\begin{align}
\frac{\mathrm{Pot}(\mathbf{R}')}{\mathrm{Pot}(\mathbf{R})} & =\frac{\Big(\sqrt{r_{i,i}^{2}+r_{i-1,i}^{2}}\Big)^{2(n-i+2)}\Big(\frac{r_{i,i}r_{i-1,i-1}}{\sqrt{r_{i,i}^{2}+r_{i-1,i}^{2}}}\Big)^{2(n-i+1)}}{r_{i-1,i-1}^{2(n-i+2)}r_{i,i}^{2(n-i+1)}}\nonumber \\
 & \leq\frac{\delta(r_{i,i}^{2}+r_{i-1,i}^{2})}{r_{i,i}^{2}+(r_{i-1,i}-\lfloor r_{i-1,i}/r_{i-1,i-1}\rceil r_{i-1,i-1})^{2}},\label{eq:pot ratio3}
\end{align}
where the last inequality comes from (\ref{eq:diagonal condition}).
Based on (\ref{eq:pot ratio3}), $\frac{\mathrm{Pot}(\mathbf{R}')}{\mathrm{Pot}(\mathbf{R})}\leq\delta$
if and only if $\lfloor r_{i-1,i}/r_{i-1,i-1}\rceil=0$. As a result,
preparing the pairs $\mathbf{t}_{i,(z_{i-1}^{*},...,z_{1}^{*})}$
and $\mathbf{r}_{i,(z_{i-1}^{*},...,z_{1}^{*})}$ based on (\ref{eq:condition1})
is only suitable for reductions before checking the DR conditions.
In case that this condition fails, we should also prepare $\mathbf{t}_{i,(z_{i-1}^{'},...,z_{1}^{'})}$
and $\mathbf{r}_{i,(z_{i-1}^{'},...,z_{1}^{'})}$ that make $\lfloor r_{i-1,i}/r_{i-1,i-1}\rceil=0$:

\begin{multline}
(z_{i-1}^{'},...,z_{1}^{'})=\arg\min_{(z_{i-1},...,z_{1})}\Big\{\left\Vert \mathbf{r}_{i,(z_{i-1},...,z_{1})}\right\Vert ,\thinspace\\
\mathrm{s.t.\thinspace}\lfloor r_{i-1,i,(z_{i-1},...,z_{1})}/r_{i-1,i-1}\rceil=0\thinspace\Big\},\label{eq:condition1-1}
\end{multline}
 in which $r_{i-1,i,(z_{i-1},...,z_{1})}$ denotes the $(i-1)$th
component of $\mathbf{r}_{i,(z_{i-1},...,z_{1})}$. In such a manner,
if a vector is swapped to the front, it is not only a short vector,
but also the one that decreases the basis potential so that this kind
of swaps cannot happen too many times.

\subsection{Algorithm description}

Combing the length reduction process above, the procedure of boosted
LLL is given in Algorithm \ref{algoLLL}. Inside the loops, it employs
a fixed structure column traverse strategy rather than using a parallel
traversing \cite{Ling2010,Zhang2012}, such that a theoretical $O(n)$
factor in bounding the number of loops can be saved. In line 4, the
PNP algorithm and rejection prepare two versions of reduced vectors.
The stronger version $\mathbf{r}_{i,(z_{i-1}^{*},...,z_{1}^{*})}$
is used before testing the DR condition (line 6), so that the new
$\mathbf{r}_{i}$ is the shortest candidate among the $L$ routes
of PNP and the old $\mathbf{r}_{i}$. If it cannot pass this test,
a weaker version $\mathbf{r}_{i,(z_{i-1}^{'},...,z_{1}^{'})}$ is
used in line 7, who has identical value in the first layer as the
Babai point and a variety of $p_{i-2}\times\cdots\times p_{1}$ routes
in the remaining layers. Lastly, line 10 restores the upper triangular
feature of $\mathbf{R}$ via a lightweight $2\times2$ Givens rotation
matrix and line 11 balances the unitary matrix. The toy example below
may help to understand our algorithm.
\begin{example}
Suppose we are reducing a matrix 
\[
\mathbf{R}=\left[\begin{array}{ccc}
1 & 0.4 & 0\\
0 & 1 & 0.52\\
0 & 0 & 1
\end{array}\right]
\]
 in round $i=3$ and executing from Line 4 of Algorithm \ref{algoLLL}.
For the PNP algorithm, we set the $L=3$ routes as $p_{2}\times p_{1}=3\times1$.
The three nearest integers to $r_{2,3}/r_{2,2}$ are $1$, $0$, $2$,
so the corresponding PNP routes are
\[
\begin{cases}
\mathbf{r}_{3,(1,1)} & =[-0.4,\thinspace-0.48,\thinspace1]^{\top},\\
\mathbf{r}_{3,(2,1)} & =[0,\thinspace0.52,\thinspace1]^{\top},\\
\mathbf{r}_{3,(3,1)} & =[-0.8,\thinspace-1.48,\thinspace1]^{\top},
\end{cases}
\]
and the rejection operation marking $\mathbf{r}_{3}$ is $\mathbf{r}_{3,(0,0)}=[0,\thinspace0.52,\thinspace1]^{\top}$.
Eq. (\ref{eq:condition1}) would choose the shortest among the above
four routes. Let it be $\mathbf{r}_{3,(2,1)}$ (or $\mathbf{r}_{3,(0,0)}$),
which is employed by Line 5. Eq. (\ref{eq:condition1-1}) can only
choose from $\mathbf{r}_{3,(1,1)}$. Then we test the DR condition
and it succeeds, so the while loop stops.
\end{example}
\begin{algorithm} \label{algoLLL} \caption{The boosted LLL algorithm.} \KwIn{original lattice basis $\mathbf{D}\in \mathbb{R}^{n\times n}$, Lovász constant $\delta$, list number $L$.}  \KwOut{reduced basis $\mathbf{D}$, unimodular matrix $\mathbf{T}$}   $[\mathbf{Q},\mathbf{R}]=\mathrm{qr}(\mathbf{D})$;   \Comment{The QR decomposition of $\mathbf{D}$}\;    $i=2$,  $\mathbf{T}=\mathbf{I}$\;  
 \While{$i\leq n$}  {  
   use (\ref{eq:condition1}) to get $[\mathbf{r}_{i,(z_{i-1}^{*},...,z_{1}^{*})},\mathbf{t}_{i,(z_{i-1}^{*},...,z_{1}^{*})}]$ and use (\ref{eq:condition1-1}) to get $[\mathbf{r}_{i,(z_{i-1}^{'},...,z_{1}^{'})},\mathbf{t}_{i,(z_{i-1}^{'},...,z_{1}^{'})}]$\;
  $\mathbf{R}_{1:n,i}=\mathbf{r}_{i,(z_{i-1}^{*},...,z_{1}^{*})}$,$\mathbf{T}_{1:n,i}= \mathbf{t}_{i,(z_{i-1}^{*},...,z_{1}^{*})}$\;   
 \If{condition (\ref{eq:diagonal condition}) fails}                           { $\mathbf{R}_{1:n,i}=\mathbf{r}_{i,(z_{i-1}^{'},...,z_{1}^{'})}$,$\mathbf{T}_{1:n,i}= \mathbf{t}_{i,(z_{i-1}^{'},...,z_{1}^{'})}$\;   
   define $\mathbf{G}$ as in (\ref{eq:given matrix})\;   
  swap $\mathbf{R}_{1:n,i}$ and $\mathbf{R}_{1:n,i-1}$,   
                    $\mathbf{T}_{1:n,i}$ and $\mathbf{T}_{1:n,i-1}$ \;       
    $\mathbf{R}_{i-1:i,1:n}\leftarrow \mathbf{G}\mathbf{R}_{i-1:i,1:n}$\;              $\mathbf{Q}_{1:n,i-1:i}\leftarrow \mathbf{Q}_{1:n,i-1:i}\mathbf{G}^\top$\;                 
              $i \leftarrow \max (i-1,2)$\;}   
                             \Else{                                                        $i \leftarrow i+1$\;}     }      
   $\mathbf{D}\leftarrow \mathbf{Q}\mathbf{R}.$          \end{algorithm}

In essence, this algorithm attempts to minimize the basis length while
keeping the Siegel condition, and the PNP algorithm offers flexible
efforts to do so. If we replace lines 4 and 5 by using the Babai point
and delete line 7, Algorithm \ref{algoLLL} degrades to the classic
LLL algorithm \cite{Lenstra1982}.

\subsection{Properties of boosted LLL}

When the boosted LLL algorithm terminates, $\delta r_{i-1,i-1}^{2}-(\frac{r_{i-1,i}}{r_{i-1,i-1}}-\lfloor\frac{r_{i-1,i}}{r_{i-1,i-1}}\rceil)^{2}r_{i-1,i-1}^{2}\leq r_{i,i}^{2}$
holds for all $2\leq i\leq n$, which ensure the Siegel properties
hold: 
\begin{equation}
|r_{i-1,i-1}|\leq\beta|r_{i,i}|.\label{eq:siegelproof}
\end{equation}
Assume the PNP algorithm has parameters $(p_{k-1},\ldots\thinspace,p_{1})$
for $2\leq k\leq n$, then $\pi_{\mathbf{R}_{\Gamma_{i}}}(\mathbf{r}_{i})$
is contained in $\mathcal{V}_{\mathbf{0}}(\mathbf{R}_{\Gamma_{i}})\cup\mathcal{P}(\left[q_{1}\mathbf{r}_{1}^{*},\ldots\thinspace,q_{i-1}\mathbf{r}_{i-1}^{*}\right])$,
where $\left\{ \mathbf{r}_{1}^{*},\ldots\thinspace,\mathbf{r}_{i-1}^{*}\right\} $
are the GSO vectors of $\mathbf{R}_{\Gamma_{i}}$. Though this region
can be much larger than $\mathcal{P}(\left[\mathbf{r}_{1}^{*},\ldots\thinspace,\mathbf{r}_{i-1}^{*}\right])$,
we have

\begin{equation}
\left\Vert \mathbf{r}_{i}\right\Vert ^{2}\leq r_{i,i}^{2}+\frac{1}{4}\sum_{j<i}r_{j,j}^{2}\label{eq:sic cond of boosted0}
\end{equation}
if $\pi_{\mathbf{R}_{\Gamma_{i}}}(\mathbf{r}_{i})\in\mathcal{P}(\left[\mathbf{r}_{1}^{*},\ldots\thinspace,\mathbf{r}_{i-1}^{*}\right])$.
If $\pi_{\mathbf{R}_{\Gamma_{i}}}(\mathbf{r}_{i})\notin\mathcal{P}(\left[\mathbf{r}_{1}^{*},\ldots\thinspace,\mathbf{r}_{i-1}^{*}\right])$,
we can always find the Babai point $\mathbf{r}_{i}'$ such that $\left\Vert \mathbf{r}_{i}\right\Vert ^{2}<\left\Vert \mathbf{r}_{i}'\right\Vert ^{2}\leq r_{i,i}^{2}+\frac{1}{4}\sum_{j<i}r_{j,j}^{2}$
due to (\ref{eq:condition1}) and (\ref{eq:condition1-1}), so condition
(\ref{eq:sic cond of boosted0}) always holds in boosted LLL.

With (\ref{eq:siegelproof}) and (\ref{eq:sic cond of boosted0}),
classical properties of LLL can be trivially proved: 

\begin{equation}
l(\mathbf{D})\leq\beta^{n-1}\lambda_{n}(\mathbf{D}),\label{eq:ild2-1}
\end{equation}
\begin{equation}
\xi(\mathbf{D})\leq\beta^{n(n-1)/2}.\label{eq:deltaILD-1}
\end{equation}

Since we have devoted much effort to implement the length reductions,
(\ref{eq:ild2-1}) and (\ref{eq:deltaILD-1}) are the least bounds
that we should expect from boosted LLL. However, moving any step forward
seems difficult because even using CVP as length reduction still fails
to generate a better explicit bound than (\ref{eq:sic cond of boosted0}).
The difficulty of improving bounds on lengths exists in all variants
of LLL, including the LLL with deep insertions (LLL-deep) \cite{Schnorr1994}.
In this regard, boosted LLL only serves as an ameliorated practical
algorithm.

\subsection{Implementation and Complexity}

The total number of loops $K$ in Algorithm \ref{algoLLL} equals
to the number testing condition (\ref{eq:diagonal condition}), whose
number of positive and negative tests are denoted as $K^{+}$ and
$K^{-}$, respectively. The total number of negative test is 
\begin{equation}
K^{-}\leq\log_{1/\delta}\frac{\mathrm{Pot}(\mathbf{D}_{0})}{\mathrm{Pot}(\mathbf{D}_{K})}=\frac{1}{\ln(1/\delta)}\times\ln\left(\frac{\mathrm{Pot}(\mathbf{D}_{0})}{\mathrm{Pot}(\mathbf{D}_{K})}\right),\label{eq:delta poly}
\end{equation}
where $\mathbf{D}_{0}$, $\mathbf{D}_{K}$ are the initial basis and
the basis after $K$ loops, respectively. In the fixed traversing
strategy, we also have $K^{+}\leq K^{-}+n-1$. We first show how to
choose $\delta$ such that the boosted LLL algorithm has the best
performance while $\frac{1}{\ln(1/\delta)}$ remains to be a polynomial
number. After that, $\frac{\mathrm{Pot}(\mathbf{D}_{0})}{\mathrm{Pot}(\mathbf{D}_{K})}$
is evaluated to complete our complexity analysis.

\subsubsection{Optimal $\delta$}

Among literature, $\delta$ is often chosen arbitrarily close to $1$
while explanations are lacking. In Micciancio's book \cite[Lem. 2.9]{Micciancio2002},
it is shown if $\delta=1/4+(3/4)^{n/(n-1)}$, then $\frac{1}{\ln(1/\delta)}\leq n^{c}$
for all $c>1$. More generally, we can define an optimal principle
of choosing $\delta$, i.e., 
\[
\delta(a^{*},n)=\frac{1}{a^{*}}+\left(\frac{a^{*}-1}{a^{*}}\right)^{n/(n-1)}
\]
where $a^{*}=\frac{1}{1-e^{-1}}$. 

With such settings, three distinctive properties exist: $\frac{1}{\ln(1/\delta)}\leq n^{c}$
for all $n$; $\delta$ is asymptotically close to $1$ so that the
algorithm has the best performance; and it is the smallest value satisfying
the previous two attributes (the fastest one among the class of best
performance). Proposition \ref{prop: delta value} justifies these
claims and the proof is given in Appendix \ref{sec: lamda1}. 
\begin{prop}
\label{prop: delta value} For arbitrary constants $a>1$, $c>1$,
if $\delta(a,\thinspace n)=\frac{1}{a}+\left(\frac{a-1}{a}\right)^{n/(n-1)}$,
then for all $n$, 
\begin{equation}
\frac{1}{\ln(1/\delta)}\leq n^{c}.\label{eq:delta1}
\end{equation}
Let $a^{*}=\lim_{n\rightarrow\infty}\arg\min_{a}\delta(a,\thinspace n)$
be defined as the universal good constant, then

\begin{equation}
a^{*}=\frac{1}{1-e^{-1}}.\label{eq:delta2}
\end{equation}

\end{prop}

\subsubsection{Total complexity in flops}

Further define $\psi(\mathbf{D})=\min_{i}r_{i,i}^{2}$ and $\Psi(\mathbf{D})=\max_{i}r_{i,i}^{2}$.
Since our length reduction does not change $r_{i,i}$ while (\ref{eq:new block})
shows that any swap can narrow the gap between $r_{i-1,i-1}$ and
$r_{i,i}$, the number of negative tests between the initial basis
$\mathbf{D}_{0}$ to the final basis $\mathbf{D}_{K}$ is 
\begin{align}
K^{-} & \leq n^{c}\ln\left(\frac{\mathrm{Pot}(\mathbf{D}_{0})}{\mathrm{Pot}(\mathbf{D}_{K})}\right)\nonumber \\
 & \leq\frac{n^{c+1}(n+1)}{2}\ln\left(\frac{\Psi(\mathbf{D}_{0})}{\psi(\mathbf{D}_{0})}\right).
\end{align}

With reference to \cite{Jalden2008}, we have $\frac{\Psi(\mathbf{D}_{0})}{\psi(\mathbf{D}_{0})}\leq\kappa(\mathbf{D}_{0})$,
where $\kappa(\mathbf{D}_{0})$ is the condition number of $\mathbf{D}_{0}$.
So if the condition number of the input basis satisfies $\ln\kappa(\mathbf{D}_{0})=O(\ln n)$,
then the number of iterations in boosted LLL is $K\leq2K^{-}+n-1=O(n^{c+2}\ln n)$,
where $c>1$ is a constant arbitrarily close to $1$. By further counting
the number of flops inside and outside the loop of Algorithm \ref{algoLLL},
the total complexity of boosted LLL is $O(Ln^{4+c}\ln n)$. 
\begin{rem}
The complexity analysis above is quite general. For instance, if $\mathbf{D}_{0}$
is Gaussian, then it follows from \cite{Jalden2008} that $\mathbb{E}(\ln\kappa(\mathbf{D}_{0}))\leq\ln n+2.24$.
In the application to IF \cite{Zhan2014IT}, we can also take a detour
to employ this property of Gaussian matrices. Firstly, the condition
number of the input basis $\mathbf{D}_{0}$ would increase if the
signal to noise ratio (SNR) rises, so it suffices to investigate the
case for infinite SNR. The target then becomes the dual of a Gaussian
random matrix that has the same condition number, so $\mathbb{E}(\ln\kappa(\mathbf{D}_{0}))\leq\ln n+2.24$
also holds in IF.
\end{rem}

\section{Application to integer forcing}

In the context of optimizing the achievable rates of IF, some results
based on LR have been presented in \cite{Sakzad2013}, where the difference
between KZ and Minkowski is not obvious because the system size is
small ($2\times2$ or $4\times4$). Since we have improved the classic
KZ and LLL, we will verify our boosted algorithms in IF by showing
their performance about ergodic rates, orthogonal defects (inversely
proportional to sum-rates), and complexity in flops.

\subsection{IF and SBP}

In this subsection, the IF transceiver architecture will be reviewed
by using real value representations for simplicity. In a MIMO system
with size $n\times n$, each antenna has a message 
\[
\mathbf{w}_{i}=[w_{i}(1),w_{i}(2),\ldots\thinspace,w_{i}(k)]^{\top},
\]
 where $i\in\left\{ 1,\ldots\thinspace,n\right\} $, $\mathbf{w}_{i}\in\mathbb{F}_{p}^{k}$,
and $\mathbb{F}_{p}$ is a finite field with size $p$. As the conversion
from message layer to physical layer, an encoder $\mathcal{E}_{i}:\thinspace\mathbb{F}_{p}^{k}\rightarrow\mathbb{R}^{n}$
maps the length-$k$ message $\mathbf{w}_{i}$ into a lattice codeword
\[
\mathbf{x}_{i}=[x_{i}(1),x_{i}(2),\ldots\thinspace,x_{i}(T)]^{\top},
\]
 where $\left\Vert \mathbf{x}_{i}\right\Vert ^{2}\leq TP$, $T$ stands
for code length and $P$ stands for SNR. All encoders operate at the
same lattice with the same rate: 
\[
R_{\mathrm{TX}}=\frac{k}{T}\log_{2}p.
\]

Let $x_{i}(j)$ be the $j$th symbol of $\mathbf{x}_{i}$, we may
write the transmitted vector across all antennas in time $j$ as $\mathbf{x}[j]=[x_{1}(j),\ldots\thinspace,x_{n}(j)]^{\top}$.
An observation $\mathbf{y}[j]\in\mathbb{R}^{n}$ can be subsequently
written as 
\begin{equation}
\mathbf{y}[j]=\mathbf{H}\mathbf{x}[j]+\mathbf{z}[j]\label{eq:if model1}
\end{equation}
 in which $\mathbf{H}\in\mathbb{R}^{n\times n}$ denotes the MIMO
channel matrix and $\mathbf{z}[j]$ is the additive white Gaussian
noise (AWGN) with $z_{i}[j]\sim\mathcal{N}(0,1)$. Let $\mathbf{Y}$,
$\mathbf{X}$, and $\mathbf{Z}$ be the concatenated $\mathbf{y}[j]$,
$\mathbf{x}[j]$ and $\mathbf{z}[j]$ from time slots $1$ to $T$.
In a linear receiver architecture, the receiver will project $\mathbf{Y}$
with a matrix $\mathbf{B}=[\mathbf{b}_{1},\ldots\thinspace,\mathbf{b}_{n}]^{\top}\in\mathbb{R}^{n\times n}$
to get the useful information $\mathbf{A}\mathbf{X}$ for further
decoding,
\begin{equation}
\mathbf{B}\mathbf{Y}=\underset{\mathrm{useful\thinspace information}}{\underbrace{\mathbf{A}\mathbf{X}}}+\underset{\mathrm{effective\thinspace noise}}{\underbrace{(\mathbf{B}\mathbf{H}-\mathbf{A})\mathbf{X}+\mathbf{B}\mathbf{Z}}}.\label{eq:effec decod}
\end{equation}
We choose $\mathbf{A}=[\mathbf{a}_{1},\ldots\thinspace,\mathbf{a}_{n}]^{\top}\in\mathbb{Z}^{n\times n}$
because these lattice codewords are closed under integer combinations.
$\mathbf{A}$ should also be full rank to avoid losing information. 

For a preprocessing matrix $\mathbf{B}$, the following computation
rate can be obtained in the $i$th effective channel if the coding
lattices satisfy goodnesses for channel coding and quantization \cite{Zhan2014IT}
\begin{equation}
R(\mathbf{H},\mathbf{a}_{i},\mathbf{b}_{i})=\frac{1}{2}\log_{2}^{+}\left(\frac{P}{\left\Vert \mathbf{b}_{i}\right\Vert ^{2}+P\left\Vert \mathbf{H}^{\top}\mathbf{b}_{i}-\mathbf{a}_{i}\right\Vert ^{2}}\right),\label{eq:rate of IF}
\end{equation}
where $\mathrm{log}_{2}^{+}(x)=\max\left\{ \log_{2}(x),0\right\} $.

The first step towards maximizing the rates is to set $\partial\left\{ \left\Vert \mathbf{b}_{i}\right\Vert ^{2}+P\left\Vert \mathbf{H}^{\top}\mathbf{b}_{i}-\mathbf{a}_{i}\right\Vert ^{2}\right\} /\partial\mathbf{b}_{i}=\mathbf{0}$
for a fixed IF coefficient matrix $\mathbf{A}$, which leads to 
\[
\mathbf{b}_{i}=(\mathbf{H}\mathbf{H}^{\top}+\frac{1}{P}\mathbf{I})^{-1}\mathbf{H}\mathbf{a}_{i}.
\]
 Plug this into (\ref{eq:rate of IF}) and use Woodbury matrix identity
for the inverse of a matrix, we have 
\begin{eqnarray}
R(\mathbf{H},\mathbf{a}_{i}) & = & \frac{1}{2}\log_{2}^{+}\left(\frac{P}{\left\Vert \mathbf{D}\mathbf{a}_{i}\right\Vert ^{2}}\right),\label{eq:rate2}
\end{eqnarray}
 where $\mathbf{D}=\Lambda^{-\frac{1}{2}}\mathbf{V}^{\top}$ and $\mathbf{V}\Lambda\mathbf{V}^{\top}=\mathbf{H}^{\top}\mathbf{H}+1/P\mathbf{I}$
is the eigendecomposition. Achieving the optimum rate is therefore
equivalent to solving SIVP on lattice $\mathcal{L}(\mathbf{D})$:
\begin{equation}
\arg\min_{\mathbf{A}\in\mathbb{Z}^{n\times n},\thinspace\mathrm{rank}(\mathbf{A})=n}\max_{i}\left\Vert \mathbf{D}\mathbf{a}_{i}\right\Vert ^{2},\label{eq:sivp}
\end{equation}
in which $\min_{\mathbf{A}\in\mathbb{Z}^{n\times n},\thinspace\mathrm{rank}(\mathbf{A})=n}\max_{i}\left\Vert \mathbf{D}\mathbf{a}_{i}\right\Vert ^{2}=\lambda_{n}(\mathbf{D})$. 

Now we explain how to obtain the estimations of messages. Upon quantizing
$\mathbf{\mathbf{B}\mathbf{Y}}$ to the fine lattice and modulo the
coarse lattice in a row-wise manner \cite{Nazer2011}, a converter
$\mathcal{D}_{i}:\thinspace\mathbb{R}^{n}\rightarrow\mathbb{F}_{p}^{k}$
then maps the physical layer codeword to a message under finite field
representations, i.e., $\hat{\mathbf{u}}_{i}=[\mathbf{W}^{\top}\mathbf{a}_{i}]\thinspace\mod\thinspace p$,
$\mathbf{W}=[\mathbf{w}_{1},\ldots\thinspace,\mathbf{w}_{n}]^{\top}$.
These combinations are then collected, so as to decode the messages
as 
\[
\left[\hat{\mathbf{w}}_{1},\ldots\thinspace,\hat{\mathbf{w}}_{n}\right]^{\top}=\mathbf{A}_{p}^{-1}\left[\hat{\mathbf{u}}_{1},\ldots\thinspace,\hat{\mathbf{u}}_{n}\right]^{\top},
\]
where $\mathbf{A}_{p}$ is a full rank matrix over $\mathbb{Z}_{p}$
and $\mathbf{A}_{p}^{-1}$ is taken over the same field.

With the above demonstrations in mind, there should be at least two
reasons for us to restrain SIVP to SBP

\begin{equation}
\arg\min_{\mathbf{A}\in\mathrm{GL}_{n}(\mathbb{Z})}\max_{i}\left\Vert \mathbf{D}\mathbf{a}_{i}\right\Vert ^{2}.\label{eq:sbp}
\end{equation}

\noindent The first reason is about flexibility. With SBP, we can
choose among lattice reduction algorithms from polynomial to exponential
complexity with guaranteed properties, and these algorithms are still
efficient when SNR is high. The second reason is about complexity,
where the inverse of $\mathbf{A}$ over finite fields is much easier
to calculate when $\mathbf{A}\in\mathrm{GL}_{n}(\mathbb{Z})$, and
algorithms for SIVP or the successive minima problem (SMP) are generally
more complicated than those of SBP \cite{Sakzad2013,Ding2015}. For
instance, we can observe that for the enumeration routines of SMP,
Minkowski reduction and boosted KZ reduction, one needs to verify
the linear independence of a new vector with previous lattice vectors
for SMP, while Minkowski reduction only needs to check the greatest
common divisor of the enumerated coefficients \cite{Zhang2012tsp}
and boosted KZ does not require such inspections.

\subsection{Simulation results}

This subsection examines the rates and complexity performance when
applying the proposed boosted KZ and boosted LLL algorithms for IF
receivers. We show the achievable rates rather than the bit error
rates of IF MIMO receivers, since the latter depend on which capacity-approaching
code for the AWGN channel is used at the transmitter. All simulations
are performed on real matrices with random entries drawn from i.i.d.
Gaussian distributions $\mathcal{N}(0,1)$. Results in the figures
are all averaged from $10^{3}$ Monte Carlo runs.

The boosted LLL algorithm is referred as ``b-LLL-$L$'' with $L$
being the total number of branches in the PNP algorithm, i.e., $L=p_{i-1}p_{i-2}\cdots p_{1}$
remains unchanged for different columns $i$'s. If $L=1$, this version
means only adding a rejection operation to the classic LLL algorithm
\cite{Lenstra1982}. When $L=3$ or $L=9$, we expand $3$ branches
in the first or first two layers of the PNP algorithm. Regarding other
typical variants of LLL, such as the effective LLL \cite{Ling2010}
and greedy LLL \cite{Wen2016}, they all boil down to the same performance
as LLL if we implement a full size reduction at the end of their algorithms,
so we omit comparing our algorithms with these variants.

The boosted KZ algorithm (``b-KZ'') is implemented as described
in Algorithm \ref{algoKZ}. To ensure a fair comparison, the KZ algorithm
follows the same routine as Algorithm \ref{algoKZ} except replacing
the ``CVP subroutine'' with a size reduction. Minkowski reduction
is also included as our reference with the label ``Minkow'', whose
implementation follows \cite[Sec. V]{Zhang2012tsp}.

\subsubsection{Achievable rate}

The actual achievable rate of the IF receivers can be quantitatively
evaluated by the ergodic rate defined by \cite{Sakzad2013}

\[
R_{E}=\mathbb{E}\Big(n\min_{i}R(\mathbf{H},\mathbf{a}_{i})\Big),
\]
where the expectation is taken over different realizations of $\mathbf{H}$,
and $R(\mathbf{H},\mathbf{a}_{i})$ was defined in (\ref{eq:rate2}). 

\begin{figure}[th]
\includegraphics[width=3.4in,height=2.6in]{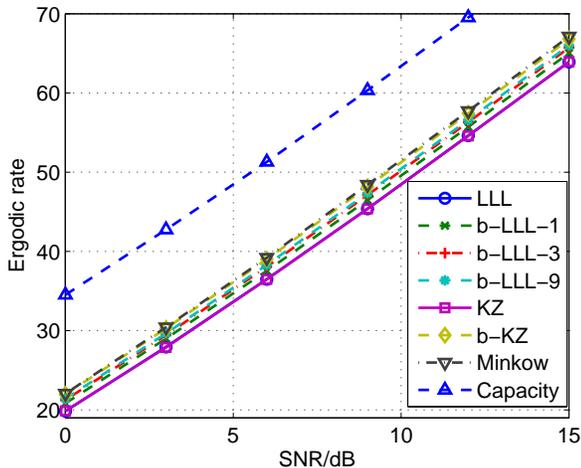}

\protect\caption{SNR versus ergodic rate for different LR algorithms.}
\label{fig rate}
\end{figure}

\begin{figure}[th]
\center

\includegraphics[width=3.4in,height=2.6in]{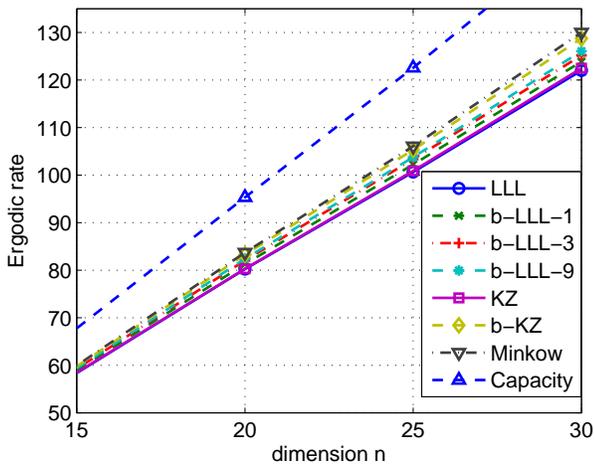}

\protect\caption{Dimension versus ergodic rate for different LR algorithms.}
\label{fig rate-1}
\end{figure}

In Fig. \ref{fig rate}, we have plotted the rate performance of different
LR algorithms in a $20\times20$ real-valued MIMO channel, in which
the channel capacity $\frac{1}{2}\log_{2}(\det(\mathbf{I}+P\mathbf{H}\mathbf{H}^{\top}))$
serves as an upper bound. In the figure, the b-LLL-1 algorithm has
higher rates than KZ and LLL, and the improvements after we increase
the list number to $L=3,\thinspace9$ can still be spotted in this
crowded figure. The b-KZ method attains almost the same rates as those
of Minkowski reduction. KZ reduction does not offer better rates than
LLL because KZ only guarantees to yield a basis with the smallest
potential, and both of them are under the curse Proposition \ref{prop:counterLLL}.

In Fig. \ref{fig rate-1}, we fixed the SNR to be $20\mathrm{dB}$
and study how the size of the system is affecting their ergodic rates.
From this graph, the differences of rates among different LR methods
amplify as dimension $n$ increases, and their mutual relations are
the same as those of Fig. \ref{fig rate}.

\subsubsection{Orthogonal defect}

The ergodic rate $R_{E}$ is only determined by the basis length $l(\mathbf{D})$.
To evaluate the sum-rates for all data streams, OD's can be employed
which are proportional to the length products of basis vectors. Such
a quantity can reveal the gaps between different algorithms more vividly.

\begin{figure}[th]
\center

\includegraphics[width=3.4in,height=2.6in]{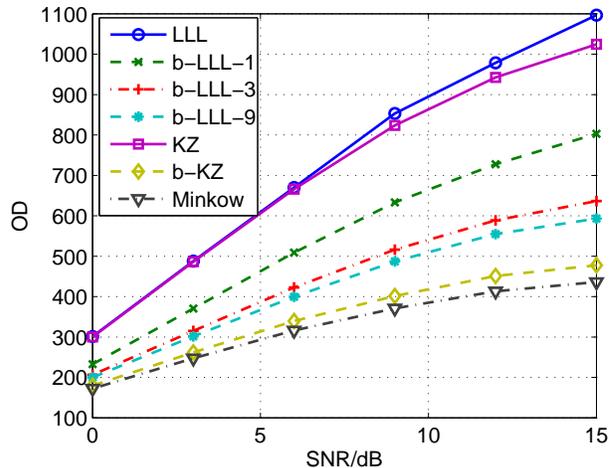}

\protect\caption{SNR versus OD for different LR algorithms.}
\label{figODrate}
\end{figure}

\begin{figure}[th]
\center

\includegraphics[width=3.4in,height=2.6in]{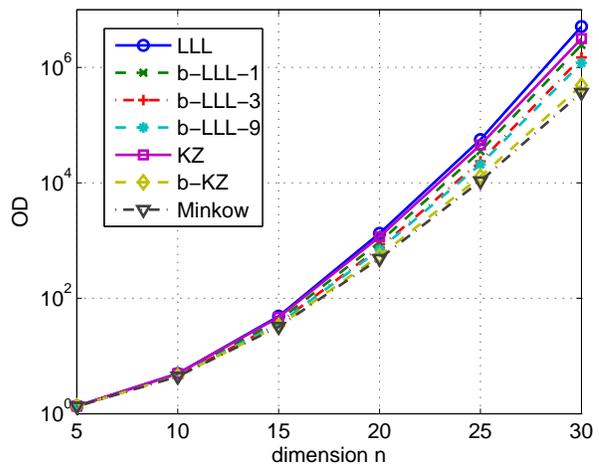}

\protect\caption{Dimension versus OD for different LR algorithms.}
\label{figODrate-1}
\end{figure}

In Figs. \ref{figODrate} (fixed size of $20\times20$) and \ref{figODrate-1}
(fixed SNR of $20\mathrm{dB}$), we have plotted the SNR versus OD,
and dimension versus OD relations for distinct lattice reduction algorithms.
From these two figures, several phenomenons can be observed. Boosted
KZ cannot surpass Minkowski reduction but remains close to it. The
performance improvements from b-LLL-$1$, to b-LLL-$3$, b-LLL-$9$
are approximately proportional to the increment in the list size $L$.
One interesting thing to observe from Fig. \ref{figODrate} is, the
performance gaps between boosted and non-boosted algorithms are becoming
larger as $P$ rises. Since $\mathbf{D}=\Lambda^{-\frac{1}{2}}\mathbf{V}^{\top}$
and $\mathbf{V}\Lambda\mathbf{V}^{\top}=\mathbf{H}^{\top}\mathbf{H}+1/P\mathbf{I}$,
the increment of $P$ is, intrinsically, changing the goodness of
the corresponding minimal basis. It also says that the possibility
of size reduction being suboptimal would increase if the lattice bases
tend to be more random. Lastly, Fig. \ref{figODrate-1} shows an evident
``Minkow < b-KZ < b-LLL-9 < b-LLL-3 < b-LLL-1 < KZ < LLL'' relation
about OD, and their OD values are much better than their theoretical
bounds (see e.g., Eqs. (\ref{eq:od bHKZ}) and (\ref{eq:deltaILD-1})).

\subsubsection{Complexity}

\begin{figure}[th]
\center

\includegraphics[width=3.4in,height=2.6in]{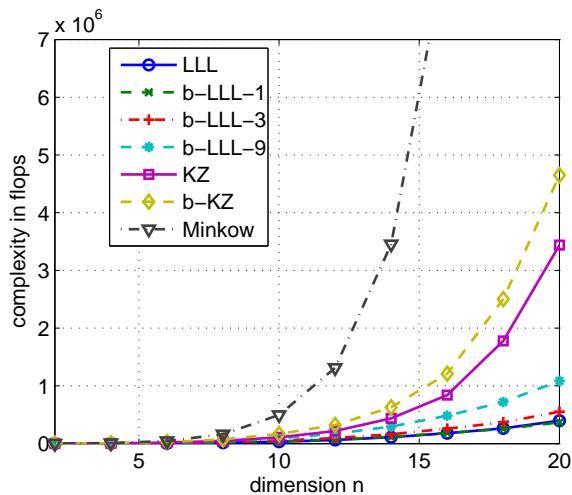}

\protect\caption{Dimension versus complexity for different LR algorithms.}
\label{fig8 complexity}
\end{figure}

In addition to our theoretical analysis on the complexity of the proposed
algorithms, we further compare their empirical costs by the expected
number of flops, which are clearly shown in Fig. \ref{fig8 complexity}.
Not surprisingly, the b-KZ algorithm spends about $1.5$ times the
efforts of KZ in the dimensions depicted in Fig. \ref{fig8 complexity},
and the b-LLL-1, 3, 9 algorithms costs around $1,\thinspace1.5,$
and $3$ times the efforts of LLL. Both b-KZ and KZ reductions have
dramatically lower number of flops than Minkowski reduction. Moreover,
the boosted LLL algorithms have much smaller complexity than KZ while
reducing the bases more effectively as Figs. \ref{fig rate} to \ref{figODrate-1}
have revealed. 

To sum up, concerning the complexity-performance tradeoffs as well
as the theoretical bounds, the boosted KZ and LLL algorithms can be
the ideal candidates for reducing lattice bases in IF with exponential
and polynomial complexity, respectively. 

\section{Open questions}

We have only demonstrated the theoretical superiority of boosted KZ
over KZ and Minkowski, while Minkowski reduction still yields shorter
vectors in our simulations. One interesting open question is whether
there exist better performance bounds for Minkowski reduction. It
is also of sufficient interests to improve the performance analysis
on boosted LLL.

\appendices{}

\section{\label{sec: short quan}Proof of proposition \ref{thm:Voroni small}}
\begin{IEEEproof}
First of all, the reducible condition $\left\Vert \mathbf{d}_{i}-\mathbf{v}\right\Vert ^{2}<\left\Vert \mathbf{d}_{i}\right\Vert ^{2}$
can be reformulated as $\left\Vert \pi_{\mathbf{D}_{\Gamma_{i}}}(\mathbf{d}_{i})-\mathbf{v}\right\Vert ^{2}<\left\Vert \pi_{\mathbf{D}_{\Gamma_{i}}}(\mathbf{d}_{i})\right\Vert ^{2}$,
which becomes equivalent to 
\begin{equation}
\left\Vert \mathbf{v}\right\Vert ^{2}/2<\langle\pi_{\mathbf{D}_{\Gamma_{i}}}(\mathbf{d}_{i}),\mathbf{v}\rangle.\label{eq:simpleproof}
\end{equation}

It is necessary to show $\langle\pi_{\mathbf{D}_{\Gamma_{i}}}(\mathbf{d}_{i}),\mathbf{v}\rangle>0$
for $\mathbf{v}\neq\mathbf{0}$ so that the inequality we pursuit
makes sense. We give a proof by contradiction. Suppose that $\theta(\pi_{\mathbf{D}_{\Gamma_{i}}}(\mathbf{d}_{i}),\mathbf{v})>\pi/2$,
due to symmetricity of the Voronoi cell $\mathcal{V}_{\mathbf{v}}(\mathbf{D}_{\Gamma_{i}})$,
there exists a symmetric point $\mathbf{d}_{i}'$ of $\pi_{\mathbf{D}_{\Gamma_{i}}}(\mathbf{d}_{i})$
on $\mathbf{v}$, such that $\theta(\mathbf{d}_{i}',\mathbf{v})>\pi/2$.
Define the half-space of $\mathbf{v}$ as $\mathcal{H}_{\mathbf{v}}=\left\{ \mathbf{x}\in\mathbb{R}^{n}\mid\langle\mathbf{v},\mathbf{x}\rangle>0\right\} $,
then the convex combination among $\left\{ \mathcal{V}_{\mathbf{v}}(\mathbf{D}_{\Gamma_{i}})\cap\mathcal{H}_{\mathbf{v}},\pi_{\mathbf{D}_{\Gamma_{i}}}(\mathbf{d}_{i}),\mathbf{d}_{i}'\right\} $
must include the origin. Then there are two lattice points ($\mathbf{0}$
and $\mathbf{v}$) inside $\mathcal{V}_{\mathbf{v}}(\mathbf{D}_{\Gamma_{i}})$,
which contradicts the basic property of a Voronoi cell, i.e., there
can be only one lattice point inside a Voronoi region. 

We proceed to prove (\ref{eq:simpleproof}). Since $\pi_{\mathbf{D}_{\Gamma_{i}}}(\mathbf{d}_{i})$
is quantized to $\mathbf{v}$, their difference $\pi_{\mathbf{D}_{\Gamma_{i}}}(\mathbf{d}_{i})-\mathbf{v}$
lies inside the Voronoi cell $\mathcal{V}_{\mathbf{0}}(\mathbf{D}_{\Gamma_{i}})$,
which yields $\langle\pi_{\mathbf{D}_{\Gamma_{i}}}(\mathbf{d}_{i})-\mathbf{v},\mathbf{w}\rangle<\left\Vert \mathbf{w}\right\Vert ^{2}/2$
for all $\mathbf{w}\in\mathcal{L}(\mathbf{D}_{\Gamma_{i}})$ and $\mathbf{w}\neq\mathbf{0}$.
As $\mathbf{v}-\pi_{\mathbf{D}_{\Gamma_{i}}}(\mathbf{d}_{i})\in\mathcal{V}_{\mathbf{0}}(\mathbf{D}_{\Gamma_{i}})$,
choose an instance of $\mathbf{w}=\mathbf{v}\neq\mathbf{0}$ for $\langle\mathbf{v}-\pi_{\mathbf{D}_{\Gamma_{i}}}(\mathbf{d}_{i}),\mathbf{w}\rangle<\left\Vert \mathbf{w}\right\Vert ^{2}/2$,
then (\ref{eq:simpleproof}) is obtained.  
\end{IEEEproof}

\section{\label{sec: kz len} Proof of proposition \ref{prop:KZ len}}
\begin{IEEEproof}
Regarding (\ref{eq:len2-1}), first recall a fact that we cannot produce
$n$ independent vectors by using a lattice of rank $n-1$, then among
$\lambda_{1}(\mathbf{D}),\ldots\thinspace,\lambda_{n}(\mathbf{D})$,
at least one of them, e.g., $\lambda_{i'}(\mathbf{D})$, corresponds
to $\mathbf{v}=x_{n}\mathbf{d}_{n}+\sum_{i=1}^{n-1}x_{i}\mathbf{d}_{i}$
where $\sum_{i=1}^{n-1}x_{i}\mathbf{d}_{i}\in\mathcal{L}(\mathbf{D}_{\Gamma_{n}})$,
$x_{n}\in\mathbb{Z}\setminus0$. With QR decomposition $[\mathbf{Q},\mathbf{R}]=\mathrm{qr}(\mathbf{D})$,
\begin{eqnarray*}
\mathbf{v} & = & \sum_{i=1}^{n}x_{i}\left(r_{i,i}\mathbf{q}_{i}+\sum_{j=1}^{i-1}r_{j,i}\mathbf{q}_{j}\right)\\
 & = & \sum_{i=1}^{n-1}x_{i}\left(r_{i,i}\mathbf{q}_{i}+\sum_{j=1}^{i-1}r_{j,i}\mathbf{q}_{j}\right)+x_{n}r_{n,n}\mathbf{q}_{n}.
\end{eqnarray*}
 Notice that $-\mathbf{v}$ also corresponds to $\lambda_{i'}(\mathbf{D})$,
so we can confine $x_{n}>0$ and consider the following two cases. 

1) If $x_{n}>1$, it is observed that  $\left\Vert \mathbf{v}\right\Vert =\left\Vert \sum_{i=1}^{n-1}x_{i}\left(r_{i,i}\mathbf{q}_{i}+\sum_{j=1}^{i-1}r_{j,i}\mathbf{q}_{j}\right)+x_{n}r_{n,n}\mathbf{q}_{n}\right\Vert \geq x_{n}|r_{n,n}|$
because the $\mathbf{q}_{i}$'s are orthogonal, which yields 
\begin{equation}
|r_{n,n}|^{2}\leq\frac{1}{x_{n}^{2}}\lambda_{i'}(\mathbf{D})^{2}\leq\frac{1}{4}\lambda_{n}(\mathbf{D})^{2}.\label{eq:len3}
\end{equation}
We then proceed to bound the last term in (\ref{eq:common equation}).
For $1\leq i\leq n$, the covering radius of lattice $\mathcal{\mathcal{L}}(\mathbf{D}_{\Gamma_{n}})$
is 
\begin{eqnarray}
\rho(\mathbf{D}_{\Gamma_{n}}) & = & \max_{\mathbf{x}}\mathrm{dist}\left(\mathbf{x},\mathcal{L}(\mathbf{D}_{\Gamma_{n}})\right)\nonumber \\
 & \leq & 1/2\sqrt{\sum_{k=1}^{n-1}r_{k,k}^{2}},\label{eq:rho gamma}
\end{eqnarray}
where the inequality is obtained after choosing $\mathbf{x}$ as a
``deep hole'' \cite[P. 33]{Conway1999} and solving this CVP by
applying Babai's nearest plane algorithm \cite{Babai1986}. Since
boosted KZ still assures $|r_{k,k}|=\lambda_{1}(\pi_{\mathbf{D}_{\Gamma_{k}}}^{\perp}([\mathbf{d}_{k},\ldots\thinspace,\mathbf{d}_{n}]))$,
and the projection of the $k$th successive minimum in $\mathbf{D}$
onto the orthogonal complement of $\mathbf{D}_{\Gamma_{k}}$ must
have a least one non-zero coefficient for $[\mathbf{d}_{k},\ldots\thinspace,\mathbf{d}_{n}]$,
we have $|r_{k,k}|\leq\lambda_{k}(\mathbf{D})$; plug this into (\ref{eq:rho gamma}),
\begin{equation}
\rho(\mathbf{D}_{\Gamma_{n}})\leq1/2\sqrt{\sum_{k=1}^{n-1}\lambda_{k}(\mathbf{D})^{2}}.\label{eq:lambdan cover}
\end{equation}
 Put (\ref{eq:len3}) and (\ref{eq:lambdan cover}) into (\ref{eq:common equation}),
then $\left\Vert \mathbf{d}_{n}\right\Vert ^{2}\leq\frac{1}{4}\lambda_{n}(\mathbf{D})^{2}+\frac{1}{4}\sum_{k=1}^{n-1}\lambda_{k}(\mathbf{D})^{2}\leq\frac{n}{4}\lambda_{n}(\mathbf{D})^{2}$.

2) If $x_{n}=1$, recall that our length reduction by CVP (line 11)
ensures $\mathbf{d}_{n}$ is the shortest vector among the set $\left\{ \mathbf{d}_{n}+\sum_{i=1}^{n-1}z_{i}\mathbf{d}_{i}\mid\forall\thinspace z_{i}\in\mathbb{Z}\right\} $,
so $\left\Vert \mathbf{d}_{n}\right\Vert \leq\lambda_{i'}(\mathbf{D})\leq\lambda_{n}(\mathbf{D})$
in such a scenario. 

Combining 1) and 2) proves (\ref{eq:len2-1}). 

As for (\ref{eq:len1}), since all sublattices $\left\{ \mathcal{L}(\mathbf{D}_{\Gamma_{i+1}})\mid1\leq i\leq n\right\} $
are also boosted KZ reduced, it follows from the proved (\ref{eq:len2-1})
that $\left\Vert \mathbf{d}_{i}\right\Vert \leq\max\left\{ 1,\thinspace\frac{\sqrt{i}}{2}\right\} \lambda_{i}(\mathbf{D}_{\Gamma_{i+1}})$.
With $|r_{i,i}|\leq\lambda_{i}(\mathbf{D})$ and the bound for the
covering radius, we also have $\left\Vert \mathbf{d}_{i}\right\Vert \leq\frac{\sqrt{i+3}}{2}\lambda_{i}(\mathbf{D})$
for all $i$. So choosing the minimum among them yields (\ref{eq:len1}).
\end{IEEEproof}

\section{\label{sec: kz gsgap}Proof of proposition \ref{prop:gs lambda1 }}
\begin{IEEEproof}
Since $|r_{i,i}|=\lambda_{1}(\mathbf{R}_{i:n,i:n})$, we apply Minkowski's
second theorem \cite[P. 202]{Cassels1971} to lattices $\mathcal{L}(\mathbf{R}_{i:n,i:n})$
with $1\leq i\leq n-1$, then we have $r_{n-j+1,n-j+1}^{2}\leq\gamma_{j}\left(\prod_{k=1}^{j}r_{n-k+1,n-k+1}^{2}\right)^{1/j}$.
As those of \cite[Prop. 4.2]{Lagarias1990}, we cancel duplicated
terms in this inequality and use induction from $\mathcal{L}(\mathbf{R}_{n-1:n,n-1:n})$
to $\mathcal{L}(\mathbf{R}_{1:n,1:n})$, then
\begin{equation}
r_{n-j+1,n-j+1}^{2}\leq\gamma_{j}\left(\prod_{k=2}^{j}\gamma_{k}^{1/(k-1)}\right)r_{n,n}^{2}.\label{eq:gs bound kz}
\end{equation}
 As $\gamma_{j}\leq\frac{2j}{3}$, we define $g(j)=\frac{2j}{3}\prod_{k=2}^{j}\left(\frac{2k}{3}\right)^{1/(k-1)}$
and evaluate this term. Let $z=k-1$, it can be shown that 
\begin{eqnarray*}
g(j) & = & \frac{8j}{9}\exp\left(\sum_{z=2}^{j-1}\ln\left(\frac{2z+2}{3}\right)\frac{1}{z}\right)\\
 & \overset{(a)}{\leq} & \frac{8j}{9}\exp\left(\sum_{z=2}^{j-1}\frac{\ln(z)}{z}\right)\\
 & \overset{(b)}{\leq} & \frac{8j}{9}\exp\left(\int_{1}^{j-1}\frac{\ln(z)}{z}\mathrm{d}z\right)\\
 & = & \frac{8j}{9}(j-1)^{\ln(j-1)/2},
\end{eqnarray*}
 where the relaxation in (a) avoids evaluating Spence's function in
the integration and Riemann integral has been used in (b). Plug this
back into (\ref{eq:gs bound kz}), we have 
\begin{equation}
r_{n-j+1,n-j+1}^{2}\leq\frac{8j}{9}(j-1)^{\ln(j-1)/2}r_{n,n}^{2},\label{eq:gs rela2}
\end{equation}
 for $2\leq j\leq n$, and this is the condition in boosted KZ that
corresponds to the Siegel condition in LLL. Let $j=n$ and apply (\ref{eq:gs rela2})
to each of $\mathcal{L}(\mathbf{R}_{1:n,1:i})$ for $2\leq i\leq n$,
then we obtain $\lambda_{1}(\mathbf{D})^{2}\leq\frac{8j}{9}(j-1)^{\ln(j-1)/2}r_{j,j}^{2}$.
By further incorporating (\ref{eq:gs rela2}) and the relation of
$\left\Vert \mathbf{d}_{i}\right\Vert ^{2}\leq r_{i,i}^{2}+\frac{1}{4}\sum_{k=1}^{i-1}r_{k,k}^{2}$,
it yields 
\begin{eqnarray*}
\left\Vert \mathbf{d}_{i}\right\Vert ^{2} & \leq & \left(1+\frac{1}{4}\sum_{j=2}^{i}\frac{8j}{9}(j-1)^{\ln(j-1)/2}\right)r_{i,i}^{2}\\
 & \leq & \left(1+\frac{2i}{9}(i-1)^{1+\ln(i-1)/2}\right)r_{i,i}^{2}
\end{eqnarray*}
 for $1\leq i\leq n$, so (\ref{eq:len gs3}) is proved.
\end{IEEEproof}

\section{\label{sec: gauss red}Proof of proposition \ref{claim: PR properties}}
\begin{IEEEproof}
It is equivalent to characterize $\mathbf{d}_{1},\thinspace\mathbf{d}_{i}$
by two numbers $u,$ $v$ in the complex plane $\mathbb{C}$, i.e.,
$u=c$ and $v=d_{1}+\sqrt{-1}d_{2}$. For an ``acute basis'' $[u,v]$
\cite[P. 76]{Nguyen2010} where $\mathcal{R}(v/u)\geq0$, the basis
reaches the first and second successive minima if and only if 
\begin{equation}
|v/u|\geq1\thinspace\thinspace\mathrm{and}\thinspace\thinspace0\leq\mathcal{R}(v/u)\leq1/2.\label{eq:req pr}
\end{equation}
All bases can be evaluated via Eq. (\ref{eq:req pr}) because either
$[u,v]$ or $[u,-v]$ must be acute. In the boosted KZ algorithm,
if we cannot reduce the length of $\mathbf{d}_{i}$ with only $\mathbf{d}_{1}$,
then $|cd_{1}|/(d_{1}^{2}+d_{2}^{2})<1/2$. In the other direction,
reducing $\mathbf{d}_{1}$ by $\mathbf{d}_{i}$ is also impossible
because $\mathbf{d}_{1}$ is already the shortest, so $|d_{1}/c|<1/2$.
By combining these two non-reducible conditions and the acute condition
of $d_{1}/c\geq0$, requirements in (\ref{eq:req pr}) can be met. 
\end{IEEEproof}

\section{\label{sec: lamda1}Proof of proposition \ref{prop: delta value}}
\begin{IEEEproof}
The proof of (\ref{eq:delta1}) follows those in \cite[Lem. 2.9]{Micciancio2002}.
To prove (\ref{eq:delta1}), it suffices to prove 
\begin{equation}
\lim_{n\rightarrow\infty}\frac{1-e^{(-(1/n)^{c})}}{\frac{a-1}{a}-(\frac{a-1}{a})^{n/(n-1)}}\leq1,\label{eq:prove lo1}
\end{equation}
 where its l.h.s. is an indeterminate form. Replace $n$ by another
variable $x$ as $n=\frac{x+1}{x}$, then by using L\textquoteright Hospital\textquoteright s
Rule, the l.h.s. of (\ref{eq:prove lo1}) becomes
\[
\lim_{x\rightarrow0}\frac{1-e^{-1/(1+1/x)^{c}}}{\frac{a-1}{a}-(\frac{a-1}{a})^{x+1}}=\lim_{x\rightarrow0}\frac{c(\frac{x}{x+2})^{c-1}\frac{1}{(x+1)^{2}}e^{-1/(1+1/x)^{c}}}{-(\frac{a-1}{a})^{x+1}\ln(\frac{a-1}{a})}=0
\]
 and thus (\ref{eq:prove lo1}) is proved. 

As for (\ref{eq:delta2}), let $\frac{\partial\delta(a,\thinspace n)}{\partial a}=\frac{n}{(n-1)a^{2}}(1-\frac{1}{a})^{1/(n-1)}-\frac{1}{a^{2}}=0$,
we obtain the stationary point of $\delta(a,\thinspace n)$ as $a'=\frac{1}{1-(1-1/n)^{n-1}}$,
where $\frac{\partial\delta(a,\thinspace n)}{\partial a}<0$ if $a\in(1,\thinspace a')$
and $\frac{\partial\delta(a,\thinspace n)}{\partial a}>0$ if $a\in(a',\thinspace\infty)$.
Notice that $(1-1/n)^{n-1}=e^{\frac{\ln(1-1/n)}{1/(n-1)}}$, then
after using L\textquoteright Hospital\textquoteright s Rule again,
we have 
\[
\lim_{n\rightarrow\infty}\frac{1}{1-(1-1/n)^{n-1}}=\lim_{n\rightarrow\infty}\frac{1}{1-e^{-1+1/n}}=\frac{1}{1-e^{-1}}.
\]

\end{IEEEproof}
\bibliographystyle{IEEEtranMine}
\phantomsection\addcontentsline{toc}{section}{\refname}\bibliography{library}

\end{document}